\documentclass{emulateapj}
\usepackage{graphicx}
\usepackage{subfigure}

\newcommand{\secpoint}{\mbox{$''\mskip-7.6mu.\,$}}
\newcommand{\secnopoint}{\mbox{$''\mskip-7.6mu\,$}}
\newcommand{\lya}{Ly$\alpha$}

\slugcomment{DRAFT: \today}

\shorttitle{NARROWBAND LYMAN-CONTINUUM IMAGING}
\shortauthors{NESTOR ET AL.}

\begin{document}

\title{Narrowband imaging of Escaping Lyman-Continuum Emission in the SSA22 Field\altaffilmark{1,2}}

\author{\sc Daniel B. Nestor, Alice E. Shapley\altaffilmark{3}}
\affil{Department of Astronomy, University of California,
Los Angeles, 430 Portola Plaza, Los Angeles, CA 90024}

\author{\sc Charles C. Steidel, Brian Siana}
\affil{California Institute of Technology, MS 105--24, 
Pasadena, CA 91125}

\altaffiltext{1}{Based, in part, on data obtained at the 
W.M. Keck Observatory, which is operated as a scientific 
partnership among the California Institute of Technology, 
the University of California, and NASA, and was made 
possible by the generous financial support of the W.M. Keck 
Foundation.}
\altaffiltext{2}{Based, in part, on data collected at the 
Subaru Telescope and obtained from the SMOKA archive, which 
is operated by the Astronomy Data Center, National 
Astronomical Observatory of Japan.}
\altaffiltext{3}{David and Lucile Packard Fellow}

\begin{abstract}
We present the results of an ultradeep, narrowband imaging
survey for Lyman-continuum (LyC) emission at $z\sim 3$
in the SSA22a field.  We employ a custom narrowband filter centered
at $\lambda=3640$~\AA\ (NB3640), which  probes
the LyC region for galaxies
at $z\geq 3.06$.  We also analyze 
new and archival 
NB4980 imaging tuned to the wavelength of the Ly$\alpha$ emission line
at $z=3.09$, and archival broadband $B$, $V$, and $R$ images
of the non-ionizing UV continuum. 
Our NB3640 images contain
26 $z\geq 3.06$ Lyman Break Galaxies (LBGs) 
as well as a set of 130 Ly$\alpha$
emitters (LAEs), identified by their excess NB4980 flux
relative to the $BV$ continuum. Six LBGs and 28 LAEs
are detected in the NB3640 image. 
LBGs appear to span a range of NB3640$-R$
colors, while LAEs appear bimodal in their NB3640$-R$ properties. 
We estimate average UV to LyC flux density ratios, 
corrected for foreground
contamination and intergalactic medium absorption, finding
$\langle F_{UV}/F_{LyC}\rangle_{corr}^{LBG} = 11.3^{+10.3}_{-5.4}$,
which implies a LBG LyC escape fraction $f_{esc}^{LyC}\sim 0.1$,
and $\langle F_{UV}/F_{LyC}\rangle_{corr}^{LAE} = 2.2^{+0.9}_{-0.6}$.
The strikingly blue 
LAE flux density ratios defy interpretation in terms of 
standard stellar population models.  
Assuming $\langle F_{UV}/F_{LyC}\rangle_{corr}^{LBG}$ applies
down to
$L=0.1 L^*$, we estimate a galaxy contribution 
to the intergalactic hydrogen
ionization rate that
is consistent with independent estimates 
based on the Ly$\alpha$ forest opacity at $z\simeq3$. 
If we assume that $\langle F_{UV}/F_{LyC}\rangle_{corr}^{LAE}$ holds at the faintest
luminosities, the galaxy contribution significantly exceeds that inferred
from the Ly$\alpha$ forest.  We interpret 
our results in terms of a model where
LyC photons escape
over only $\sim 10-20$\% of solid angle.
When advantageously oriented, 
a galaxy will exhibit a low UV to LyC ratio, an
effect enhanced for more compact galaxies. This model, however, does not adequately
explain the extremely blue NB3640$-R$ colors measured for some LAEs in
our sample. Further follow-up study of these faint LAEs is crucial, given
the potentially important contribution similar objects make to the process
of reionization.

\end{abstract}


\section{Introduction}    \label{sect:intro}

The thermal state of the intergalactic medium (IGM) is governed by
its spatial density distribution, the spectral shape and intensity of the metagalactic
ionizing background, and the distribution of the sources of the
background.  Identifying these sources constitutes a critical step towards
understanding the physics and thermal history of the IGM.  QSOs are 
efficient producers of far-UV radiation and dominate the Lyman-continuum (LyC)
emissivity that maintains the ionization state of the IGM up to the
peak of AGN activity at redshift $z\sim 2$ \citep{cowie2009}.  Beyond $z\sim2$, however,
their contribution decreases drastically with their decreasing number
density \citep{hopkins2007}.  As it has been demonstrated that
reionization was completed by $z\ga 6$ \citep{fan2006,becker2007}, it is
generally assumed that the ionizing flux required for 
reionization was produced in star-forming galaxies.  Direct
measurements of ionizing radiation escaping from galaxies at $z\ga6$
are not possible because of the large optical depth to
LyC photons in the high-redshift IGM.  However, the measurement can in
principle be made at lower redshift where the IGM is less opaque.

Although
observations of star-forming galaxies at relatively low redshift ($0.5
\la z\la 2$) have failed to detect escaping LyC photons
\citep[e.g.,][]{malkan2003,siana2007,grimes2009,siana2010,bridge2010}, there has
been much recent progress in the search for LyC 
emission at higher ($z \sim 3$ - 4) redshift.  Building on the initial detection
by \citet{steidel2001} in a composite spectrum of 29 LBGs at $\langle z\rangle = 3.4$,
\citet{shapley2006} presented
the first spectroscopic detections of LyC emission
from individual objects. In a sample of
14 objects with deep Keck/LRIS spectra,
two exhibited significant LyC
emission, implying $f^{LyC}_{esc}>0.15$.
Deep Keck/LRIS spectroscopy for a $z\sim 3$ sample larger by an
order of magnitude is forthcoming, and can
be used to generalize the initial spectroscopic
results (Steidel  et al., in preparation).

An alternative technique for studying the escape of LyC
emission consists of narrowband imaging.
Measuring the density of LyC radiation with imaging requires a filter that only transmits
flux below the Lyman limit, and also one that is relatively narrow.
A filter covering too broad a range below the Lyman limit will indicate more
about the statistics of IGM absorption than the absorption
intrinsic to galaxies. Specifically, a filter width comparable to or larger than
than the wavelength range corresponding to one LyC
photon mean free path will typically contain significant absorption
due to intervening Lyman limit systems, thus making it more difficult
to infer the amount of escaping ionizing radiation in the immediate
vicinity of a high-redshift galaxy. Current estimates of the mean
free path at $z\sim 3$ are $\sim 70-85$~Mpc (proper), assuming
the current concordance cosmology \citep{faucher2008,prochaska2009,songaila2010}.
This range in mean free path corresponds to a rest-frame wavelength interval
of $\sim 850-910$~\AA. Since standard broadband $U$ filters probe a larger
wavelength baseline than this, it is advantageous to adopt a custom
setup, including a narrowband filter whose transmission curve probes
within the rest wavelength range corresponding to one mean free path.

Key for maximizing the efficiency of narrowband imaging in a fixed
rest-frame wavelength range is the existence of a large
sample of galaxies in a single field at roughly the same redshift.
In this respect, the SSA22a field serves as an ideal target for
narrowband imaging searches for LyC emission at $z\sim 3$.
This field has been shown to contain a significant overdensity
of Lyman Break Galaxies (LBGs) at $z=3.09\pm 0.03$ \citep{steidel1998}.
In addition to the 27 LBGs spectroscopically confirmed to lie within
the $z=3.09$ redshift spike, narrowband imaging
tuned to the wavelength of Ly$\alpha$ emission at $z=3.09$ has
been used to identify large samples of Ly$\alpha$ emitters (LAEs)
and extended Ly$\alpha$ ``blobs"
at the same redshift \citep{steidel2000, matsuda2004,hayashino2004}.

Escaping LyC emission at $z\geq 3$ detected by narrowband imaging was
first reported by \citet{iwata2009}.
Using Subaru/Suprime-Cam and a ``NB359" filter, with a central
wavelength 3590\AA, and FWHM of 150\AA, these authors
searched for LyC emission for sources at $z\geq 3.06$,
and reported detections for 7 LBGs and 10 LAEs.
In some cases
there are spatial offsets of several kpc between
the centroids of ionizing and non-ionizing UV emission,
which may distinguish among different models for the
escape of LyC emission \citep[e.g.,][]{gnedin2008}. Furthermore,
in some of the fainter LAEs, the high apparent ratio
of escaping ionizing to non-ionizing UV radiation
challenges standard models for the intrinsic spectral
energy distribution of star-forming galaxies \citep{inoue2010a,inoue2010b}.
In a parallel and independent experiment, we have used
Keck/LRIS and a narrowband ``NB3640" filter, similar to the
Suprime-Cam NB359 filter, in order to collect even deeper
images of the LyC region for $z\sim 3$ galaxies in the SSA22a
field. We have also obtained significantly deeper imaging
tuned to the wavelength of Ly$\alpha$, enabling more robust
identification of LAEs, down to fainter flux limits.

In this paper, we present the results of our narrowband
imaging survey for LyC emission at $z\sim 3$ in the SSA22a field.
\S\ref{sect:data} describes the observations and
data reduction. In \S3, we present narrow and broadband photometric
measurements, the identification of a sample of LAEs
based on our new, deeper data set, and the method used
for matching apparent LyC detections with
known $z\sim 3$ targets.  We also discuss
potential emission contamination along the line of
sight in our narrowband imaging.  \S4 contains
our results, including estimates of the true
ratios of ionizing to non-ionizing flux densities for our
targets, the corresponding escape fractions, and the implied space density
of ionizing radiation.  In \S5, we discuss the cosmological
implications of these results, and offer
a concluding summary in \S6.  Throughout the paper we employ the AB
magnitude system, quote 
flux and luminosity densities in units of $F_{\nu}$ and $L_{\nu}$, respectively, 
and assume a cosmology with $\Omega_m=0.3$, $\Omega_{\Lambda}=0.7$,
and $H_0= 70$~km~s$^{-1}$~Mpc$^{-1}$.

\section{Observations and Data Reduction}    \label{sect:data}

\begin{figure*}
\epsscale{0.95}
\plotone{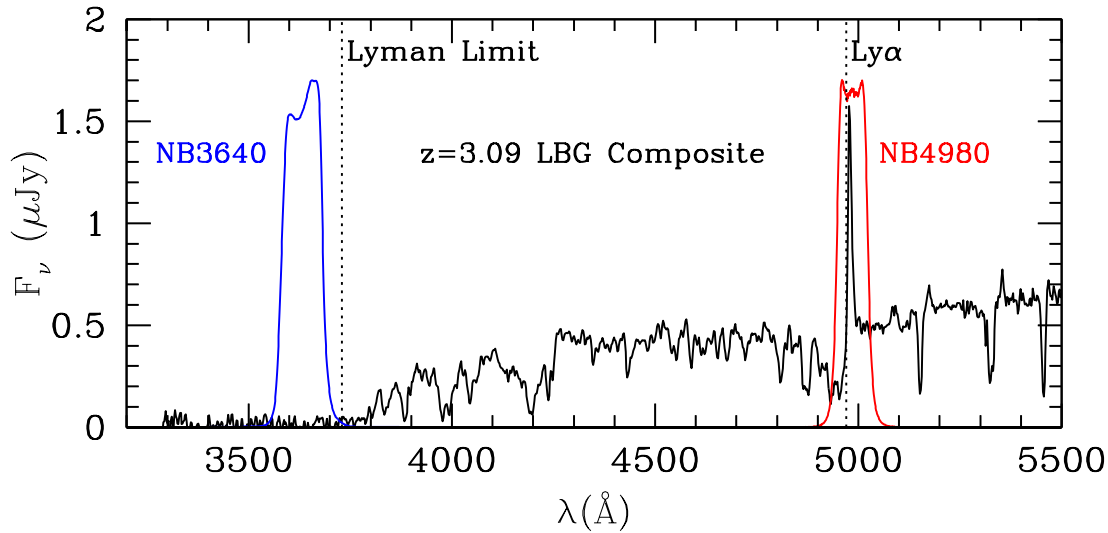}
\vspace{0.0cm}
\caption{
\small
Narrowband filter transmission curves overlaid on an LBG
composite spectrum from \citet{shapley2006}, with $\langle z
\rangle=3.09$.
The NB3640 filter probes the LyC region for galaxies at $z\geq 3.06$,
while the NB4980 filter contains the Ly$\alpha$ feature
for galaxies at $z\sim 3.09$, the redshift of the significant
galaxy overdensity in the SSA22a field. Dotted lines mark both the Lyman
limit
and Ly$\alpha$ feature at $z=3.09$.
\label{fig:nbfilter_plot}
}
\end{figure*}

A custom narrowband filter was used for LyC observations,
with central wavelength 3635 \AA\ and FWHM 
100 \AA, hereafter referred to as NB3640. The filter was 
originally designed to detect Ly$\alpha$ emission at $z=2$ 
(Andrew Blain, private communication), but its 
specifications are such that no significant contamination 
($\leq 1$\%) longwards of the Lyman limit (912\AA) will 
enter the filter for redshifts $z\geq 3.06$.  Furthermore, the narrow NB3640 bandpass probes the wavelength region well
within one mean free path below the Lyman limit at the redshift for the
majority of our target galaxies
\citep{faucher2008,prochaska2009,songaila2010}, thus minimizing the
correction for intergalactic medium absorption (IGM) that is required to
infer $f^{LyC}_{esc}$ \citep{shapley2006}.
Figure~\ref{fig:nbfilter_plot} shows the transmission curve 
of the NB3640 filter overlaid on a composite $z=3.09$ LBG 
spectrum from \citet{shapley2006}. Also shown in 
Figure~\ref{fig:nbfilter_plot} is the transmission curve of 
an additional custom narrowband filter tuned to the 
wavelength of Ly$\alpha$ at $z=3.09$, the SSA22a spike 
redshift. This filter has central wavelength 4985 \AA\ and 
FWHM 80 \AA, hereafter referred to as NB4980.

We obtained NB3640 imaging data using the blue side of the 
Low Resolution Imaging Spectrometer 
\citep[LRIS;][]{steidel2004}. The NB3640 imaging data were 
collected on 6--7 August 2005 and 2--4 June 2008 (UT), for a 
total of 13.2 hours of integration. During the August 2005 
run, the red side of LRIS was simultaneously used to collect 
NB4980 data, for a total integration time of 9.4 hours. The 
blue side (LRIS-B) detector has a plate scale of 
0\secpoint135, while the red side (LRIS-R) has 0\secpoint211 
pixels.  The ``d460" dichroic beam splitter was used to send 
light bluer than $\sim 4600$~\AA\ towards the LRIS-B, while 
longer wavelengths were directed towards LRIS-R.  Conditions were 
photometric during both observing runs, with seeing in the 
final NB3640 and NB4980 stacked images of 0\secpoint80. The 
pointing center of the LRIS imaging field was 
$\alpha=$22:17:28.8, $\delta=$+00:14:36, and the area of the 
final trimmed stacks is $5.5'\times 7.6'$. Contained within 
this footprint are 26 of the 33 LBGs in SSA22a with 
spectroscopic redshifts $z\geq 3.06$, such that the NB3640 
probes bluewards of the Lyman limit. There are also two 
$z>3.06$ QSOs, and 41 LBG photometric candidates without 
spectroscopic redshifts. The images were taken at a sky 
position angle of $\theta=180^{\circ}$, with the long axis of the image
oriented along the North-South direction.  The LRIS-B 
4K~$\times$~4K CCD mosaic detector consists of two 
2K~$\times$~4K chips with slightly different ($\sim20$\%) quantum 
efficiency at 3640\AA, and the pointing center and sky 
position angle were chosen to maximize the number of 
$z\geq3.06$ targets falling on the blue side chip with 
better quantum efficiency.

Individual narrowband exposures were reduced using standard 
procedures of flatfielding, background subtraction, and 
removal of cosmic rays. For both NB3640 and NB4980 imaging, 
flatfields were constructed from de-registered images of the 
twilight sky. For imaging in each narrowband filter, tasks 
from the IRAF ``images" and ``mscred'' packages were used to 
solve for the world coordinate system of each exposure, 
sample each exposure onto a reference world coordinate 
system, and shift and combine all individual exposures into 
a final stacked image. During combination of individual 
exposures, which ranged in airmass from $1.06-1.37$, scaling 
was applied so that each exposure effectively was observed 
at the minimum airmass. Furthermore, LRIS-B images were 
resampled onto the 0\secpoint211 LRIS-R pixel scale. 
Narrowband images were calibrated onto the AB system using 
observations of spectrophotometric standard stars from the 
list of \citet{massey1988}. NB3640 and NB4980 AB magnitudes 
were also corrected for a Galactic extinction of $E(B-V)=0.08$, 
based on $IRAS$ $100\mu\mbox{m}$ cirrus emission maps 
\citep{schlegel1998}.

We augmented the LRIS data set with both narrowband and 
broadband archival imaging data taken with Suprime-Cam 
\citep{miyazaki2002} on the Subaru Telescope. Additional 
data include narrowband images taken through a 
filter with central wavelength 4977 \AA\ and FWHM 80 \AA, 
i.e. almost identical to the one we used for Keck/LRIS 
observations, and described previously in 
\citet{matsuda2004}, \citet{hayashino2004}, and 
\citet{iwata2009}. These ``NB497'' images were obtained on 
8--9 September 2002 (UT), for a total of 7.2 hours of 
integration time, and an image quality in the stack of 
0\secpoint76 FWHM. Broadband $B$- and $V$-band images were 
obtained on 21 September 2003 (UT) (PI: Hu) with integration 
times of 1.1 and 0.8 hours, respectively, and respective 
image quality of 0\secpoint76 and 0\secpoint60. Finally, we 
analyze $R$-band imaging from 20 October 2001 (UT) 
\citep{hu2004}, with an integration time of 0.8 hours and 
seeing of 0\secpoint56. All Suprime-Cam data were retrieved 
from the Subaru archive, SMOKA \citep{baba2002}, and reduced 
with the SDFRED reduction package 
\citep{yagi2002,ouchi2004}. Subaru $B$, $V$-band, and $R$ 
images were flux calibrated using a combination of 
spectrophotometric and Landolt \citep{landolt1992} standard 
stars. A summary of both Keck/LRIS and Subaru/Suprime-Cam 
observations is presented in Table~\ref{tab:obs}.

\begin{deluxetable*}{cccc} 
\tablewidth{0pt} 
\footnotesize
\tablecaption{Description of Observations \label{tab:obs}}
\tablehead{
\colhead{Filter} &
\colhead{Telescope/Instrument} &
\colhead{Seeing FWHM (\secnopoint) } &
\colhead{Exposure (s)} 
}
\startdata 
NB3640  & Keck/LRIS-B       & 0.80 & 47636  \\
NB4980  & Keck/LRIS-R       & 0.80 & 33880  \\
        & Subaru/Suprime-Cam & 0.76 & 25800  \\
&&&\\
\hline
&&&\\

B       & Subaru/Suprime-Cam & 0.76 & ~3927  \\
V       & Subaru/Suprime-Cam & 0.60 & ~3000  \\
R       & Subaru/Suprime-Cam & 0.56 & ~3000  \\
\enddata
\end{deluxetable*}

Given the similarity of the Ly$\alpha$ narrowband filters 
used for LRIS and Suprime-Cam imaging, in terms of central 
wavelength and FWHM, we combined NB4980 and NB497 images, 
scaling them to a common zeropoint, and weighting them 
according to their respective depths. This combined NB4980 
image is $\sim 0.4$~mag deeper than the Subaru-only image 
presented in \citet{matsuda2004}. The selection of 
Ly$\alpha$ emitters (LAEs) requires a continuum image 
probing the same effective wavelength as the narrowband. 
Since the combination of $B$ and $V$ filters straddles 
the NB4980 bandpass, a linear combination of the two filters 
suffices for the matched continuum image. We constructed a 
$BV$ continuum image by first smoothing the $V$-band image 
so that the $B$ and $V$ image PSFs were matched, and then 
scaling the PSF-matched images to the same zeropoint. We 
then executed: $BV = 0.44 B + 0.56 V$, with the weights 
determined by the relative wavelengths of the broadband and 
NB4980 filters. We note that the linear combination of $B$ 
and $V$ images described here differs from that in 
\citet{matsuda2004} (i.e., $(2B+V)/3$), and better reflects 
the relative wavelengths probed by the $B$, $V$, and NB4980 
filters. Finally, we smoothed the $BV$ continuum image so that 
the PSF matched that of the $NB4980$ combined image. As 
described in \S\ref{s:laec}, selection of LAEs is based on the 
NB4980 magnitudes and $BV-NB4980$ colors of objects. 
However, for examining the spatial distribution of 
Ly$\alpha$ emission (and absorption), isolated from the 
local continuum, we also constructed a continuum-subtracted 
Ly$\alpha$ image, hereafter referred to as $LyA$.

To estimate the noise properties and depth of the images
from which photometry was measured (NB3640, LRIS+Suprime-Cam
NB4980, $BV$-continuum, and $R$), we evaluated the counts in apertures
distributed randomly in regions free of detected sources. To facilitate 
comparison with \citet{iwata2009}, we used 1\secpoint2 diameter circular apertures.
The standard deviation of the counts in empty apertures yields the associated
1$\sigma$ magnitude limit, from which an estimate of the 3$\sigma$ limiting
magnitude can be derived. We find  3$\sigma$ limiting magnitudes of
27.90, 27.90, 28.40, and 27.60, respectively,
for the NB3640, combined LRIS+Suprime-Cam NB4980, $BV$-continuum,
and $R$ images. Our NB3640 image is therefore roughly 0.6 deeper
than the analogous NB359 image presented in \citet{iwata2009}.
As described in \S\ref{s:phot}, this simple estimate of the noise
properties of our images does not take into account the full
systematic uncertainty associated with source detection and evaluation. Monte
Carlo simulations replicating our detection and measurement
process with artificial galaxies of known magnitude
indicate larger photometric errors than
inferred from the stated depths above. All error bars
presented in later sections are based on the more conservative Monte Carlo
simulations.

Finally, for detailed rest-frame UV (non-ionizing)
continuum morphologies of our SSA22a targets, we assembled archival {\it Hubble
Space Telescope} ({\it HST}) Advanced Camera for Surveys (ACS) imaging of
the field, taken with the F814W filter.
These data include 3 pointings from the Gemini 
Deep Deep Survey \citep[][P.I.D. 9760]{abraham2007}, 
with 5-orbit exposure times, and 3 pointings from the 
program described in \citet{geach2009} (P.I.D. 10405),
with 3-orbit exposure times. Of the 34 targets with NB3640 detections
described in \S\ref{s:dets}, 25 have deep ACS coverage.

\section{Analysis}
\subsection{Photometric measurements and uncertainties}
\label{s:phot}
Source identification and photometry were performed using SExtractor
version 2.5.0 \citep{bertin96}.  We first smoothed the NB3640
image using a circular Gaussian kernel with $\sigma = 1$ pixel $=$
0\secpoint211.  We then ran SExtractor in ``dual-image mode'', 
using the smoothed image to detect sources though the measurements were
performed on the unsmoothed image.  We found this method produced
superior results in terms of detecting faint sources while not producing
overly-large apertures, compared to the use of SExtractor's built-in filtering option.
Similarly, the NB4980 image was used to identify sources and define
isophotes for measurements in the $BV$ image for the purpose of
identifying LAE candidates (see below).

Due to the details of the radiative transfer and sources of the
corresponding photons, any detected \lya\ emission and escaping LyC flux will not
necessarily be co-spatial with either each other or with the bulk of the
rest-frame UV flux in a given galaxy.  Therefore, single-image mode was employed 
for measurements in the NB4980 and $R$ images with no prior used to
define apertures, and we report ``Kron-like''  (i.e., SExtractor
MAG\_AUTO) total magnitudes for these and the NB3640 images.  All
reported colors are determined by taking the
differences of these magnitudes.  This method allows more flexibilty (e.g.,
in avoiding contamination from neighboring sources) and decreases the
photometric uncertainty, as compared to using apertures of fixed shape and
size large enough to capture all of the flux from the most extended
sources.  Furthermore, we restrict ``detections'' to magnitude ranges
for which any systemics in the photometry are small compared to the
random error (see below).

In addition to the estimates of the depth reached in each image
described above (\S\ref{sect:data}), we ran simulations of our method in order to
obtain a detailed description of the combined systematic and statistical
uncertainties in our measurements.  We placed one hundred 
simulated sources of known magnitude into the NB3640 image, the $R$ image,
and of known magnitude and color simultaneously into the NB4980 and
$BV$ images.  The simulated sources were placed at random positions
chosen to avoid both real and previously placed simulated sources.
SExtractor was then run on these simulated source-added images in the
same manner as above (i.e., in single-image mode on $R$ and NB4980
and dual-image mode for detection/measurement on smoothed/unsmoothed
NB3640 and NB4980/$BV$) 
and magnitudes were recorded for all recovered simulated sources.
This process was repeated until a large enough total number (typically $\sim
50000$) of simulated sources had been recovered to populate each bin
in magnitude (and color) to a statistically significant degree.  The average
differences and standard
deviations of the recovered from input magnitudes then define any
systematics and the uncertainty as a function of
magnitude (and color).  We list the systematics and uncertainties in
NB3640 and $R$ in Table~\ref{t:errs}, and present those for NB4980 and
$BV$-NB4980 in Table~\ref{t:errs2}. 

\begin{deluxetable}{ccccc}
\tablewidth{0pt} 
\footnotesize
\tablecaption{Uncertainties in simulated photometry: $R$ and NB3640. \label{t:errs}}
\tablehead{
\colhead{magnitude range \tablenotemark{a}} & \colhead{$\Delta R$ \tablenotemark{b}} & \colhead{$\sigma_{R}$ \tablenotemark{c}} & \colhead{$\Delta$NB3640 \tablenotemark{b}} & \colhead{$\sigma_{\mathrm{NB3640}}$ \tablenotemark{c}}
}
\startdata
23.0 - 23.5 & 0.00 & 0.05 & $-0.01$ & 0.03 \\
23.5 - 24.0 & 0.00 & 0.08 & $-0.01$ & 0.04 \\
24.0 - 24.5 & $-0.01$ & 0.11 & $-0.02$ & 0.06 \\
24.5 - 25.0 & $-0.01$ & 0.16 & $-0.03$ & 0.10 \\
25.0 - 25.5 & $-0.02$ & 0.23 & $-0.04$ & 0.15 \\
25.5 - 26.0 & $-0.01$ & 0.29 & $-0.06$ & 0.20 \\
26.0 - 26.5 & $0.01$ & 0.37 & $-0.06$ & 0.29 \\
26.5 - 27.0 & $0.13$ & 0.40 & $-0.03$ & 0.38 \\
27.0 - 27.5 & $0.43$ & 0.55 & $0.03$ & 0.49 \\
\enddata
\tablenotetext{a}{Recovered range of $R$ or NB3640 magnitude for simulated galaxies.}
\tablenotetext{b}{Average value of the difference between measured and input magnitudes.  Significant departures from zero imply systematic biases in the photometry.}
\tablenotetext{c}{The standard deviation in the difference between measured and input magnitudes.}
\end{deluxetable}

\begin{deluxetable*}{lccccc}
\tablewidth{0pt} 
\tabletypesize{\scriptsize}
\tablecaption{Uncertainties in simulated photometry: NB4980 and $BV-$NB4980. \label{t:errs2}}
\tablehead{
\colhead{NB4980 \tablenotemark{a}} & \colhead{$BV-$NB4980 \tablenotemark{a}} & \colhead{$\Delta$NB4980 \tablenotemark{b}} & \colhead{$\sigma_{\mathrm{NB4980}}$ \tablenotemark{c}} & \colhead{$\Delta$($BV-$NB4980) \tablenotemark{b}} & \colhead{$\sigma_{BV-\mathrm{NB3640}}$ \tablenotemark{c}}
}
\startdata
23.0 - 23.5........ & 0.0 - 0.2 & 0.01 & 0.03 & 0.00 & 0.03 \\ 
            & 0.2 - 0.4 & 0.01 & 0.04 & 0.00 & 0.03 \\ 
            & 0.4 - 0.6 & 0.01 & 0.04 & 0.00 & 0.03 \\ 
            & 0.6 - 0.8 & 0.01 & 0.03 & -0.00 & 0.03 \\ 
            & 0.8 - 1.0 & 0.01 & 0.04 & -0.00 & 0.04 \\ 
            & 1.0 - 1.2 & 0.01 & 0.04 & -0.00 & 0.05 \\ 
            & 1.2 - 1.4 & 0.01 & 0.03 & -0.00 & 0.06 \\ 
            & 1.4 - 1.6 & 0.01 & 0.03 & -0.01 & 0.07 \\ 
            & 1.6 - 1.8 & 0.01 & 0.04 & -0.01 & 0.09 \\ 
            & 1.8 - 2.0 & 0.01 & 0.03 & -0.02 & 0.11\smallskip\\ 
23.5 - 24.0........ & 0.0 - 0.2 & 0.01 & 0.05 & 0.01 & 0.04 \\ 
            & 0.2 - 0.4 & 0.01 & 0.05 & 0.00 & 0.04 \\ 
            & 0.4 - 0.6 & 0.02 & 0.05 & 0.00 & 0.04 \\ 
            & 0.6 - 0.8 & 0.01 & 0.05 & 0.00 & 0.05 \\ 
            & 0.8 - 1.0 & 0.01 & 0.06 & 0.00 & 0.06 \\ 
            & 1.0 - 1.2 & 0.01 & 0.05 & -0.01 & 0.07 \\ 
            & 1.2 - 1.4 & 0.01 & 0.05 & -0.01 & 0.09 \\ 
            & 1.4 - 1.6 & 0.01 & 0.05 & -0.01 & 0.09 \\ 
            & 1.6 - 1.8 & 0.01 & 0.05 & -0.03 & 0.13 \\ 
            & 1.8 - 2.0 & 0.02 & 0.05 & -0.03 & 0.15\smallskip\\ 
24.0 - 24.5........ & 0.0 - 0.2 & 0.02 & 0.08 & 0.00 & 0.05 \\ 
            & 0.2 - 0.4 & 0.01 & 0.08 & 0.01 & 0.05 \\ 
            & 0.4 - 0.6 & 0.02 & 0.09 & 0.01 & 0.06 \\ 
            & 0.6 - 0.8 & 0.02 & 0.08 & 0.00 & 0.06 \\ 
            & 0.8 - 1.0 & 0.02 & 0.06 & -0.00 & 0.07 \\ 
            & 1.0 - 1.2 & 0.02 & 0.07 & -0.01 & 0.09 \\ 
            & 1.2 - 1.4 & 0.01 & 0.08 & -0.01 & 0.11 \\ 
            & 1.4 - 1.6 & 0.01 & 0.08 & -0.02 & 0.14 \\ 
            & 1.6 - 1.8 & 0.02 & 0.08 & -0.04 & 0.18 \\ 
            & 1.8 - 2.0 & 0.01 & 0.08 & -0.02 & 0.18\smallskip\\ 
24.5 - 25.0........ & 0.0 - 0.2 & 0.02 & 0.10 & 0.02 & 0.07 \\ 
            & 0.2 - 0.4 & 0.03 & 0.12 & 0.01 & 0.08 \\ 
            & 0.4 - 0.6 & 0.03 & 0.12 & 0.01 & 0.08 \\ 
            & 0.6 - 0.8 & 0.03 & 0.11 & 0.00 & 0.09 \\ 
            & 0.8 - 1.0 & 0.02 & 0.12 & 0.01 & 0.11 \\ 
            & 1.0 - 1.2 & 0.02 & 0.11 & -0.02 & 0.14 \\ 
            & 1.2 - 1.4 & 0.02 & 0.13 & -0.01 & 0.16 \\ 
            & 1.4 - 1.6 & 0.02 & 0.13 & -0.02 & 0.20 \\ 
            & 1.6 - 1.8 & 0.03 & 0.11 & -0.04 & 0.21 \\ 
            & 1.8 - 2.0 & 0.02 & 0.12 & -0.07 & 0.27\smallskip\\ 
25.0 - 25.5........ & 0.0 - 0.2 & 0.04 & 0.15 & 0.03 & 0.10 \\ 
            & 0.2 - 0.4 & 0.03 & 0.17 & 0.03 & 0.10 \\ 
            & 0.4 - 0.6 & 0.02 & 0.18 & 0.02 & 0.11 \\ 
            & 0.6 - 0.8 & 0.02 & 0.16 & 0.01 & 0.13 \\ 
            & 0.8 - 1.0 & 0.03 & 0.14 & 0.00 & 0.14 \\ 
            & 1.0 - 1.2 & 0.03 & 0.16 & -0.01 & 0.19 \\ 
            & 1.2 - 1.4 & 0.01 & 0.18 & -0.05 & 0.24 \\ 
            & 1.4 - 1.6 & 0.03 & 0.16 & -0.05 & 0.26 \\ 
            & 1.6 - 1.8 & 0.02 & 0.17 & -0.08 & 0.32 \\ 
            & 1.8 - 2.0 & 0.03 & 0.16 & -0.10 & 0.38\smallskip\\ 
25.5 - 26.0........ & 0.0 - 0.2 & 0.04 & 0.22 & 0.05 & 0.13 \\ 
            & 0.2 - 0.4 & 0.06 & 0.21 & 0.03 & 0.13 \\ 
            & 0.4 - 0.6 & 0.04 & 0.22 & 0.04 & 0.15 \\ 
            & 0.6 - 0.8 & 0.05 & 0.23 & 0.01 & 0.19 \\ 
            & 0.8 - 1.0 & 0.01 & 0.25 & -0.01 & 0.22 \\ 
            & 1.0 - 1.2 & 0.02 & 0.24 & -0.03 & 0.24 \\ 
            & 1.2 - 1.4 & 0.05 & 0.24 & -0.05 & 0.28 \\ 
            & 1.4 - 1.6 & 0.03 & 0.24 & -0.15 & 0.44 \\ 
            & 1.6 - 1.8 & 0.02 & 0.20 & -0.11 & 0.45 \\ 
            & 1.8 - 2.0 & 0.05 & 0.21 & -0.13 & 0.43\smallskip\\ 
26.0 - 26.5........ & 0.0 - 0.2 & 0.07 & 0.26 & 0.04 & 0.15 \\ 
            & 0.2 - 0.4 & 0.10 & 0.28 & 0.05 & 0.17 \\ 
            & 0.4 - 0.6 & 0.07 & 0.26 & 0.04 & 0.20 \\ 
            & 0.6 - 0.8 & 0.08 & 0.28 & 0.03 & 0.22 \\ 
            & 0.8 - 1.0 & 0.05 & 0.28 & 0.00 & 0.27 \\ 
            & 1.0 - 1.2 & 0.07 & 0.33 & -0.02 & 0.33 \\ 
            & 1.2 - 1.4 & 0.08 & 0.29 & -0.07 & 0.37 \\ 
            & 1.4 - 1.6 & 0.04 & 0.26 & -0.14 & 0.48 \\ 
            & 1.6 - 1.8 & 0.06 & 0.27 & -0.12 & 0.48 \\ 
            & 1.8 - 2.0 & 0.06 & 0.28 & -0.12 & 0.49 \\ 
\enddata
\tablenotetext{a}{Recovered magnitude or color range of simulated galaxies.}
\tablenotetext{b}{Average value of the difference between measured and input magnitudes or colors.  Significant departures from zero imply systematic biases in the photometry.}
\tablenotetext{c}{The standard deviation in the difference between measured and input magnitudes or colors.}
\end{deluxetable*}

\subsection{\lya\ emitter candidates}
\label{s:laec} 

\begin{figure*}
\epsscale{1.0}
\plotone{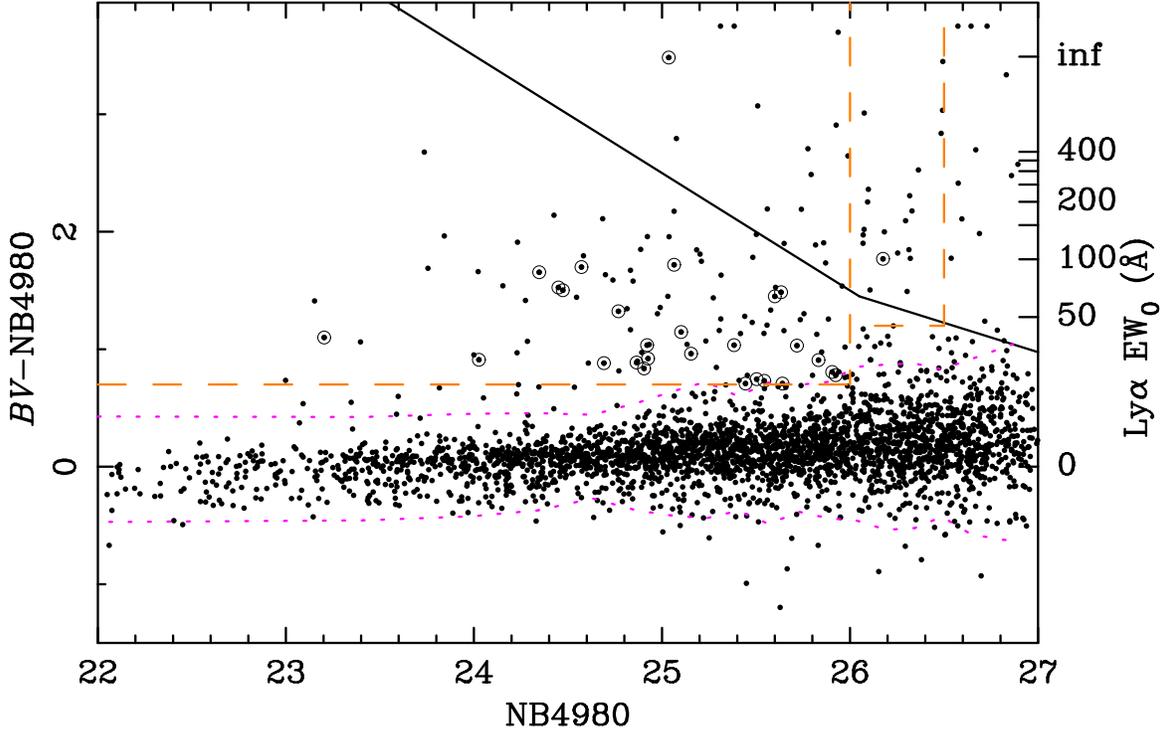}
\caption{\small 
Color-magnitude diagram for sources detected in
  the NB4980 image.  Right axis indicates the rest-frame 
  Ly$\alpha$ equivalent width (assuming $z=3.09$) implied by the $BV-$NB4980 color for
  sources at $3.054 \la z \la 3.12$.
  Isophotes for $BV-$NB4980 colors are defined in the
  NB4980 image, while NB4980 magnitudes are ``Kron-like'' (i.e., SExtractor
  MAG\_AUTO) total magnitudes.  The solid curve indicates our formal lower-limit
  threshold in $BV-$NB4980, though sources with measured values above
  this limit are plotted with their measured values, excepting those
  with measured colors $BV-\mathrm{NB4980}>3.75$ which are plotted as $BV-\mathrm{NB4980}=3.75$.
  Dotted lines indicate $\pm 3 \sigma$ scatter in the data and
  dashed lines indicate our limits defining the LAE
  candidate sample.  Circled points indicate sources with NB3640
  detections. \label{f:cmd1}}
\end{figure*}

The wavelength of redshifted \lya\ falls within our NB4980 filter for sources at
$3.054 \lesssim z \lesssim 3.120$.\footnote{Our NB3640 filter is
opaque to non-ionizing UV flux above $z\ga3.06$ and transparent to only negligible
levels of non-ionizing flux at $z \simeq 3.054$.}
Figure~\ref{f:cmd1} shows isophotal $BV-$NB4980 color as a function of total
NB4980 magnitude for all sources detected in the NB4980
image.  As the isophotes defined in the NB4980 image were used for the
$BV$ photometry, source-matching was not an issue.  Note that
objects identified by SExtractor within \lya\ ``blobs'' \citep[e.g.,][]{steidel2000}
were removed from the catalog.  The solid line indicates our formal
lower-limit in $BV-$NB4980 color.  For sources bright in NB4980, the
limit in color depends primarily on the detection limit $BV=27.5$.
At fainter narrowband magnitudes, the color limit is
determined empirically from the Monte Carlo simulations described above. 
 
Following
\citet{steidel2000}, we identify as LAE candidates 
sources having $BV-$NB4980 $\ge 0.7$.  The dotted lines in Figure~\ref{f:cmd1} 
indicate the $\pm 3\sigma$ scatter of the main distribution of
sources, which is dominated by objects outside of the redshift range
$3.054 \lesssim z \lesssim 3.120$.  The right-hand axis of Figure~\ref{f:cmd1} shows
the \lya\ rest-frame equivalent width, $EW_0$, implied by the $BV-$NB4980 color for
sources in this redshift range.  
The limit $BV$$-$NB4980 $\ge 0.7$ (corresponding to an observed \lya\
equivalent width $\simeq80$\AA) is above this scatter down to 
NB4980 $\simeq 26$.  A total of 110 sources meet the criteria
$BV$$-$NB4980 $\ge 0.7$ and NB4980 $\le 26$, which
we use to define our main LAE candidate sample.  For reference, our
simulations indicate the uncertainty in color is
$\sigma_{BV-\mathrm{NB4980}} \simeq 0.2$ for $BV$$-$NB4980 $=
0.7$ and NB4980 $= 26$.  We also consider sources with $26 <$NB4980 $\le 26.5$
with the more conservative color-cut of $BV-$NB4980 $\ge 1.2$ as an
additional ``faint''  LAE sample.  Twenty sources meet these
criteria.  As these sources are selected using different criteria,
they are excluded from any following calculations using the ensemble
of LAEs, unless otherwise noted.
The limits defining these sample are shown in Figure~\ref{f:cmd1} with dashed
lines.  

To test the robustness of the LAE selection technique, we compared
the equivalent widths implied by the $BV-$NB4980 colors 
for the 18 LBGs in the field with $3.05 \le z \le 3.12$ (excluding three within \lya\ ``blobs'')
to those determined from existing LRIS spectra
\citep{shapley2003,shapley2006}.  The equivalent width values
determined from the two methods correlate strongly.  The
photometrically determined widths are on average larger, which 
may be due to slit-losses in the spectra as the spatial extent of the \lya\ 
emission tends to be larger than that of the rest-frame UV
\citep{hayashino2004,steidel2011}.
Importantly, the spectra of all five LBGs selected as LAEs exhibit
\lya\ in emission, and no LBGs with spectroscopic observed equivalent widths
greater than 80\AA\ are missed.  Additionally, we verified that none of the 45
galaxies with spectroscopic redshifts $z<3.05$ in the field were
selected as LAEs.  We also ran SExtractor on the $LyA$ (NB4980$-BV$
difference) image; all of the 130 LAE candidates were recovered as
detections.

\subsection{Object matching}
\label{s:match}
For each LBG, we compiled the positions and photometric measurements for
any SExtractor detection in the NB3640 image within a 2\arcsec\ (corresponding to 
15~kpc at $z=3.09$) offset from the center of the $R$-band aperture.
For each LAE candidate (hereafter, LAE), we retained any detections in the
NB3640 and $R$ image within 2\arcsec\ of the NB4980 aperture center.
We visually inspected each matched source in all available bands,
including the deep $BV$ and the high image quality $R$, 
and removed any obvious false matches corresponding to neighboring sources.
In rare cases, the deblending produced by SExtractor was
unsatisfactory and the photometry was recomputed using the IRAF task
{\it phot}.  

Six of the 26 LBGs were detected in our NB3640 image.  We retain the
NB3640 detection near C16 despite the large 
(1\secpoint9) offset, as there are no neighbors detected in any of
the other bands coincident with the position of the NB3640 flux.
Of the 110 main
sample LAEs, 27 were detected in NB3640 and 72 were detected in $R$
with $R\le27$, which we adopt as our formal detection limit based on our
simulations discussed in \S\ref{s:phot}.  As with C16, a small number
of detections near LAEs having fairly large ($\sim$ 1\secpoint5 -
2\arcsec) offsets are retained as the LAE is the closest source 
detected in any of the other bands.  We account for the possibility of
mismatched NB3640 detections in \S\ref{s:contam}.
One of the 20 faint sample
LAEs was also detected in the NB3640 image, and two detected in the
$R$-band image with $R\le27$.


\subsection{LBGs and LAEs with NB3640 detection}
\label{s:dets}

Table~\ref{t:lbgs} lists
the coordinates, redshifts based on \lya\ emission and interstellar
absorption (when available), $R$ and NB3640 magnitudes, and
spatial offsets of the $R$ and NB3640 emission centroids ($\Delta_R$) for the six LBGs with NB3640 detections.
Of the two LBGs reported by \citet{shapley2006} to exhibit LyC flux in
their spectra, we detect one (C49), consistent with the results of
\citet{iwata2009}. 
The 28 (27 main and one faint sample) LAEs with NB3640 detections are marked with
circles in Figure~\ref{f:cmd1}, and the coordinates, photometry,
Ly$\alpha$ $EW_0$, and $\Delta_R$ and
$LyA$-NB3640 ($\Delta_{LyA}$) spatial offsets
are presented in Table~\ref{t:laes}.  We also note that five of the 26 LBGs in our sample 
qualify as LAEs by our criteria, though none of these are
detected in NB3640.  Information for LBGs and LAEs with no NB3640
detection is given in the appendix.  

\begin{deluxetable*}{cccccccccc}
\tablewidth{0pt} 
\tabletypesize{\scriptsize}
\tablecaption{Photometry for LBGs with NB3640 detections. \label{t:lbgs}}
\tablehead{
\colhead{ID} & \colhead{RA} & \colhead{Dec} & \colhead{$z_{em}$ \tablenotemark{a}} & \colhead{$z_{abs}$ \tablenotemark{b}} & \colhead{$R$} & \colhead{NB3640} & \colhead{$\Delta_R$ \tablenotemark{c}} & \colhead{$\frac{F_{UV}}{F_{LyC}}_{obs}$ \tablenotemark{d}} & \colhead{$\frac{F_{UV}}{F_{LyC}}_{cor}$ \tablenotemark{e}} \\
 & \colhead{(J2000)} & \colhead{(J2000)} & & & & & & & 
}
\startdata
MD46 & 22:17:27.28 & 0:18:09.7 & 3.091 & 3.080 & 23.49 & 25.22 & 1\secpoint0 & $4.9 \pm 0.7$ & $2.4 \pm 1.0$ \\
C16 & 22:17:31.95 & 0:13:16.3 & \nodata & 3.0651 & 23.62 & 26.43 & 1\secpoint9 & $13.3 \pm 4.0$ & $6.8 \pm 3.3 $ \\
C49 & 22:17:19.81 & 0:18:18.8 & 3.1629 & 3.1492 & 23.81 & 26.84 & 0\secpoint4 & $16.4 \pm 6.1$ & $5.9 \pm 4.1 $ \\
D17 & 22:17:18.86 & 0:18:17.0 & 3.0898 & 3.0697 & 24.29 & 27.00 & 0\secpoint9 & $12.2 \pm 5.1$ & $5.9 \pm 3.5 $ \\
aug96M16 & 22:17:30.86 & 0:13:10.8 & 3.298 & 3.285 & 24.47 & 25.23 & 0\secpoint7 & $2.0 \pm 0.4$ & $0.2 \pm 0.4 $ \\
MD32 & 22:17:23.70 & 0:16:01.6 & 3.102 & \nodata & 25.14 & 25.51 & 0\secpoint4 & $1.4 \pm 0.4$ & $0.6 \pm 0.3 $ \\
\enddata
\tablenotetext{a}{\mbox{Ly$\alpha$} emission redshift.}
\tablenotetext{b}{Instellar absorption redshift.}
\tablenotetext{c}{Spatial offset between the centroids of $R$-band and NB3640 emission.}
\tablenotetext{d}{Observed ratio and uncertainty in non-ionizing UV and LyC emission, inferred from the NB3640$-R$ color.  This value has not been corrected for either contamination by foreground sources or IGM absorption.}
\tablenotetext{e}{Ratio of non-ionizing UV continuum and LyC emission, corrected for both foreground contamination and IGM absorption.}
\end{deluxetable*}

\begin{deluxetable*}{ccccccccccccc}
\tablewidth{0pt} 
\tabletypesize{\scriptsize}
\tablecaption{Photometry for LAEs with NB3640 detections. \label{t:laes}}
\tablehead{
\colhead{ID} & \colhead{RA} & \colhead{Dec} & \colhead{4980} & \colhead{$BV-$NB4980} & \colhead{$EW_0$ \tablenotemark{a}} & \colhead{$R$} & \colhead{NB3640} & \colhead{$\Delta_R$ \tablenotemark{b}} & \colhead{$\Delta_{LyA}$ \tablenotemark{b}} & \colhead{$\frac{F_{UV}}{F_{LyC}}_{obs}$ \tablenotemark{c}} & \colhead{$\frac{F_{UV}}{F_{LyC}}_{cor}$ \tablenotemark{d}} \\ 
 & \colhead{(J2000)} & \colhead{(J2000)} & & & \colhead{(\AA)} & & & & & & 
}
\startdata
003 & 22:17:24.79 & 0:17:17.4 & 23.20 & 1.10 & 39 & 24.42 & 24.74 & 0\secpoint3 & 0\secpoint6	 &  $1.3 \pm 0.1$ &	   $0.6 \pm 0.3 $ \\
010 & 22:17:20.38 & 0:18:04.2 & 24.03 & 0.91 & 28 & 25.77 & 26.74 & 0\secpoint3 & 0\secpoint3	 &  $2.4 \pm 1.1$ &	   $1.2 \pm 0.7 $ \\
016 & 22:17:35.61 & 0:18:00.2 & 24.35 & 1.66 & 86 & 26.24 & 26.91 & 0\secpoint9 & 0\secpoint8	 &  $1.8 \pm 0.9$ &	   $0.9 \pm 0.6 $ \\
018 & 22:17:39.01 & 0:17:26.4 & 24.45 & 1.53 & 72 &  26.25 & 25.69 & 0\secpoint1 & 1\secpoint0	 & $0.6 \pm 0.2$ &	   $0.3 \pm 0.2 $ \\
019 & 22:17:26.15 & 0:13:20.1 & 24.47 & 1.50 & 70 & 25.70 & 26.24 & 0\secpoint4 & 0\secpoint9	 &  $1.6 \pm 0.6$ &	   $0.8 \pm 0.4 $ \\
021 & 22:17:18.77 & 0:15:18.1 & 24.57 & 1.70 & 92 & $>$27 & 27.17 & 1\secpoint3 & 1\secpoint4	 & $< 1.8$ &	   $< 0.9$ \\
025 & 22:17:36.74 & 0:16:28.8 & 24.69 & 0.88 & 27 & 25.54 & 25.85 & 0\secpoint3 & 1\secpoint2	 &  $1.3 \pm 0.4$ &	   $0.6 \pm 0.3$ \\
028 & 22:17:31.80 & 0:17:17.9 & 24.77 & 1.32 & 54 & 25.50 & 26.71 & 0\secpoint3 & 0\secpoint9	 & $3.1 \pm 1.3$ &	   $1.5 \pm 0.9 $ \\
034 & 22:17:23.41 & 0:16:35.4 & 24.86 & 0.89 & 27 & 25.42 & 25.76 & 0\secpoint5 & 0\secpoint0	 &  $1.4 \pm 0.4$ &	   $0.7 \pm 0.3 $ \\
038 & 22:17:34.77 & 0:15:41.3 & 24.90 & 0.84 & 25 &  26.17 & 25.82 & 0\secpoint1 & 0\secpoint7	 & $0.7 \pm 0.3$ &	   $0.3 \pm 0.2 $ \\
039 & 22:17:24.08 & 0:11:31.7 & 24.92 & 1.03 & 35 & 26.48 & 26.77 & 0\secpoint4 & 0\secpoint8	 & $1.3 \pm 0.7$ &	   $0.6 \pm 0.4 $ \\
041 & 22:17:24.54 & 0:15:06.7 & 24.93 & 0.92 & 29 & 25.97 & 25.94 & 0\secpoint4 & 0\secpoint5	 & $1.0 \pm 0.4$ &	   $0.5 \pm 0.3 $ \\
046 & 22:17:21.47 & 0:14:54.6 & 25.04 & $>2.54$ & $>$277 & $>$27 & 26.43 & 1\secpoint8 & 1\secpoint8	    & $< 0.8$ &    $< 0.4$ \\
048 & 22:17:27.37 & 0:16:51.5 & 25.06 & 1.72 & 94 & 26.73 & 26.00 & 2\secpoint2 & 1\secpoint7	 & $0.5 \pm 0.2$ &	   $0.2 \pm 0.2 $ \\
051 & 22:17:33.72 & 0:15:04.9 & 25.10 & 1.15 & 41 & 26.26 & 27.21 & 0\secpoint3 & 0\secpoint7	 &  $2.4 \pm 1.3$ &	   $1.2 \pm 0.8 $ \\
053 & 22:17:34.70 & 0:16:33.4 & 25.15 & 0.96 & 31 & 26.53 & 26.98 & 0\secpoint6 & 0\secpoint9	 &  $1.5 \pm 0.8$ &	   $0.7 \pm 0.5 $ \\
064 & 22:17:35.42 & 0:12:14.6 & 25.38 & 1.04 & 35 & 26.05 & 26.74 & 0\secpoint1 & 1\secpoint0	 &  $1.9 \pm 0.9$ &	   $0.9 \pm 0.6 $ \\
069 & 22:17:18.96 & 0:11:12.0 & 25.44 & 0.71 & 19 & 24.62 & 27.22 & 0\secpoint2 & 0\secpoint9	 &  $10.9 \pm 5.1$ &	   $5.3 \pm 3.3 $ \\
074 & 22:17:36.47 & 0:12:54.8 & 25.50 & 0.75 & 21 & 26.16 & 25.52 & 0\secpoint1 & 0\secpoint7	 & $0.6 \pm 0.2$ &	   $0.3 \pm 0.1 $ \\
077 & 22:17:37.95 & 0:11:01.3 & 25.54 & 0.73 & 20 & 26.01 & 26.36 & 0\secpoint2 & 1\secpoint1	 &  $1.4 \pm 0.6$ &	   $0.7 \pm 0.4 $ \\
081 & 22:17:29.22 & 0:14:48.7 & 25.60 & 1.45 & 65 & $>$27 & 26.79 & 0\secpoint6\tablenotemark{e} & 0\secpoint7	   & $< 1.2$ &     $< 0.6$ \\
083 & 22:17:28.46 & 0:12:08.9 & 25.63 & 1.48 & 68 &  26.46 & 26.84 & 0\secpoint4 & 0\secpoint6	 &  $1.4 \pm 0.7$ &	   $0.7 \pm 0.5 $ \\
084 & 22:17:19.90 & 0:15:14.9 & 25.64 & 0.71 & 20 & $>$27 & 26.50 & 0\secpoint1 & 0\secpoint8	 & $< 0.9$ &	   $< 0.4$ \\
087 & 22:17:37.07 & 0:13:21.5 & 25.72 & 1.03 & 35 & $>$27 & 27.26 & 0\secpoint1\tablenotemark{e} & 1\secpoint6   & $< 2.0$ &     $< 1.0$ \\
096 & 22:17:38.93 & 0:11:37.4 & 25.83 & 0.91 & 28 & 26.64 & 26.45 & 0\secpoint6 & 0\secpoint6	 & $0.8 \pm 0.4$ &	   $0.4 \pm 0.3 $ \\
101 & 22:17:25.33 & 0:17:22.5 & 25.90 & 0.81 & 24 & 26.89 & 26.58 & 0\secpoint2 & 1\secpoint4	 & $0.8 \pm 0.4$ &	   $0.4 \pm 0.2 $ \\
102 & 22:17:24.00 & 0:16:27.6 & 25.92 & 0.78 & 22 & 25.91 & 26.05 & 0\secpoint1 & 0\secpoint2	 &  $1.1 \pm 0.4$ &	   $0.6 \pm 0.3 $ \\
118 & 22:17:35.28 & 0:10:59.9 & 26.18 & $>1.39$ & $>$243 & 26.20 & 26.11 & 0\secpoint4 & 1\secpoint4       & $0.9 \pm 0.4$ & 	 $0.4 \pm 0.3 $ \\
\enddata
\tablenotetext{a}{\mbox{Ly$\alpha$} rest equivalent width estimated from $BV-$NB4980 color.}
\tablenotetext{b}{Spatial offset between the centroids of $R$ or Ly$\alpha$ and NB3640 emission.}
\tablenotetext{c}{Observed ratio and uncertainty in non-ionizing UV and LyC emission, inferred from the NB3640$-R$ color.  This value has not been corrected for either contamination by foreground sources or IGM absorption.}
\tablenotetext{d}{Ratio of non-ionizing UV continuum and LyC emission, corrected for both foreground contamination and IGM absorption.}
\tablenotetext{e}{Offset determined from centroid of BV detection.}
\end{deluxetable*}

\begin{figure*}
\epsscale{.65}
\plotone{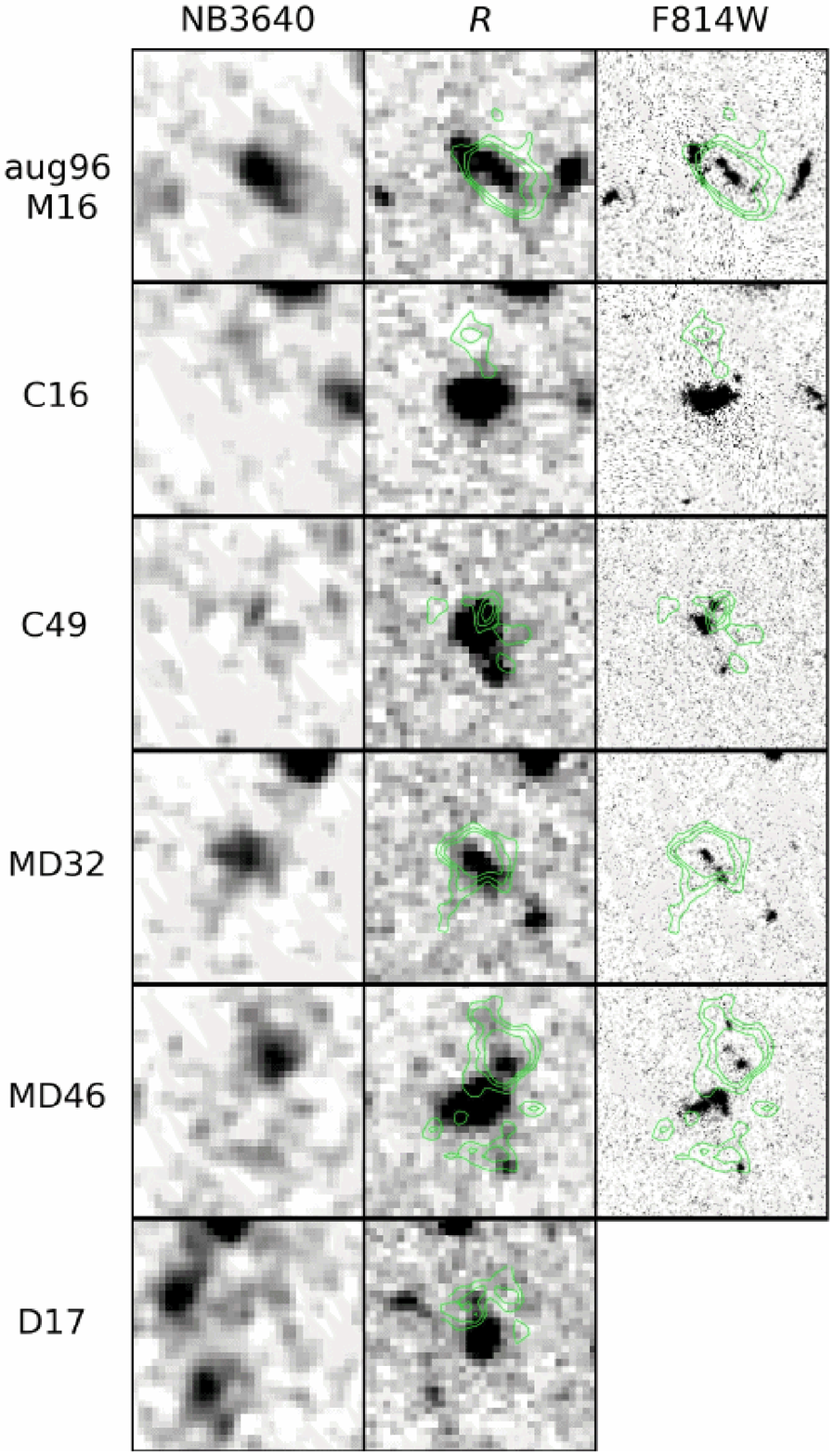}
\caption{\small 
LBGs with $z>3.06$ detected in NB3640.  Images are centered on the
  $R$-band centroid and span 7\arcsec $\times
  7$\arcsec.  The NB3640 filter (image shown here after smoothing) 
  samples the rest-frame LyC, while the $R$ and F814W
  filters sample the rest-frame non-ionizing UV continuum.  \label{f:lbgs}}
\end{figure*}

\begin{figure*}
\epsscale{.826}
\plotone{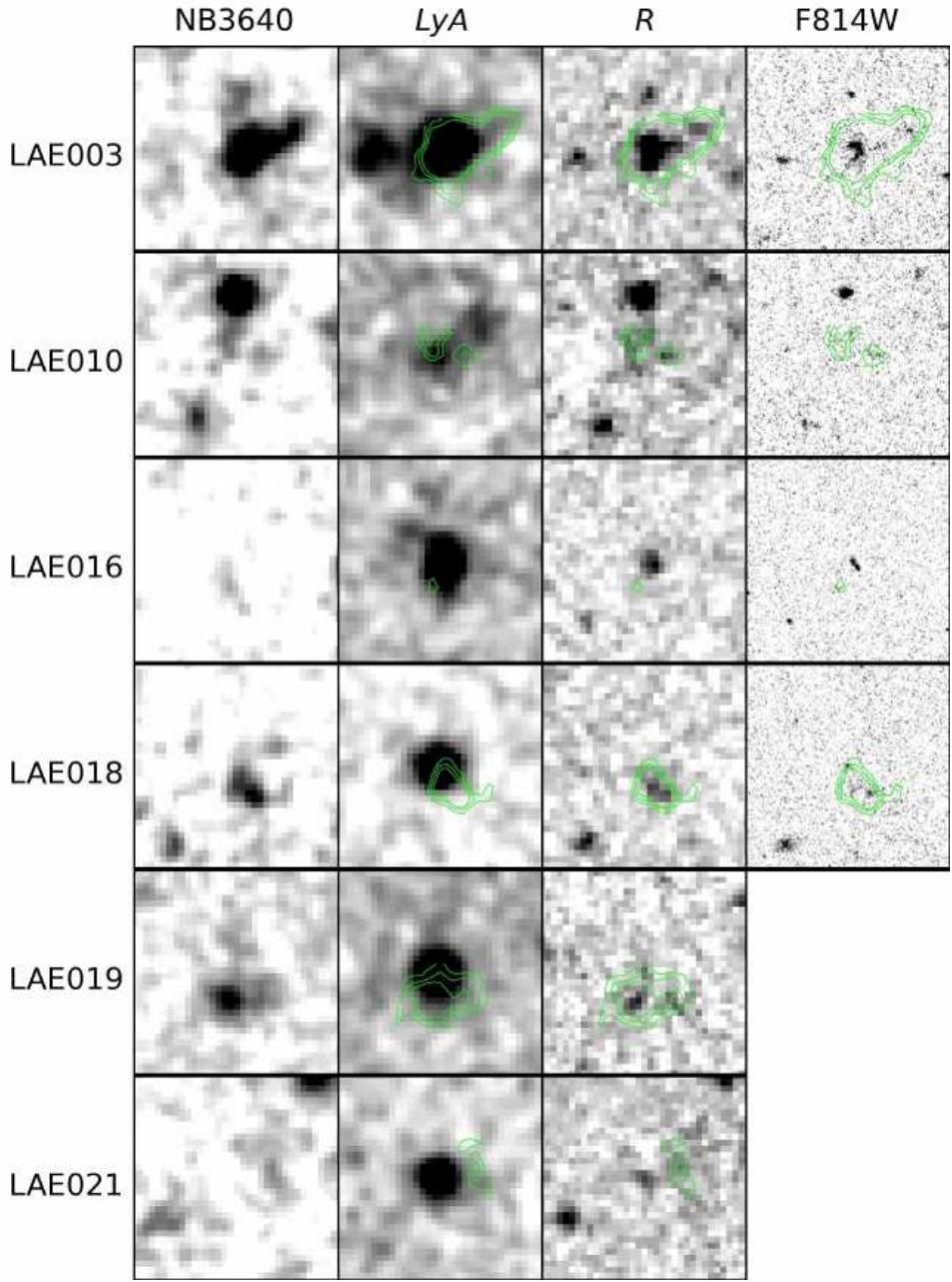}
\caption{\small 
As in Figure~\ref{f:lbgs}, but for LAEs with NB3640 detections.  Also
shown is the $LyA$ (i.e., $BV$-continuum subtracted NB4980, see text) image.  Images
are centered on the $LyA$ detection and span 7\arcsec $\times
  7$\arcsec.  \label{f:laes1}}
\end{figure*}

\begin{figure*}
\epsscale{.826}
\plotone{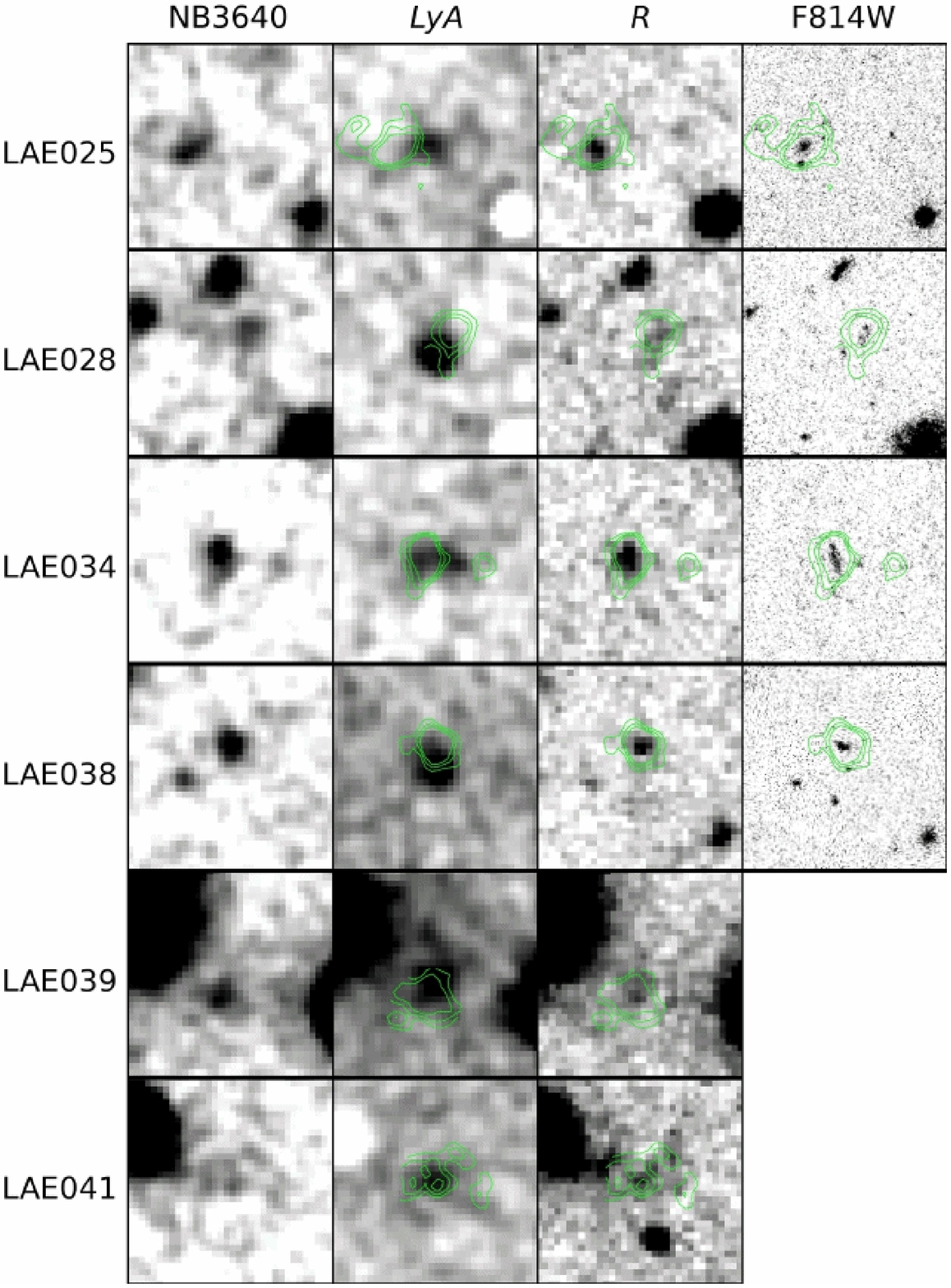}
\caption{\small 
Same as Figure~\ref{f:laes1}. \label{f:laes2}}
\end{figure*}

\begin{figure*}
\epsscale{.826}
\plotone{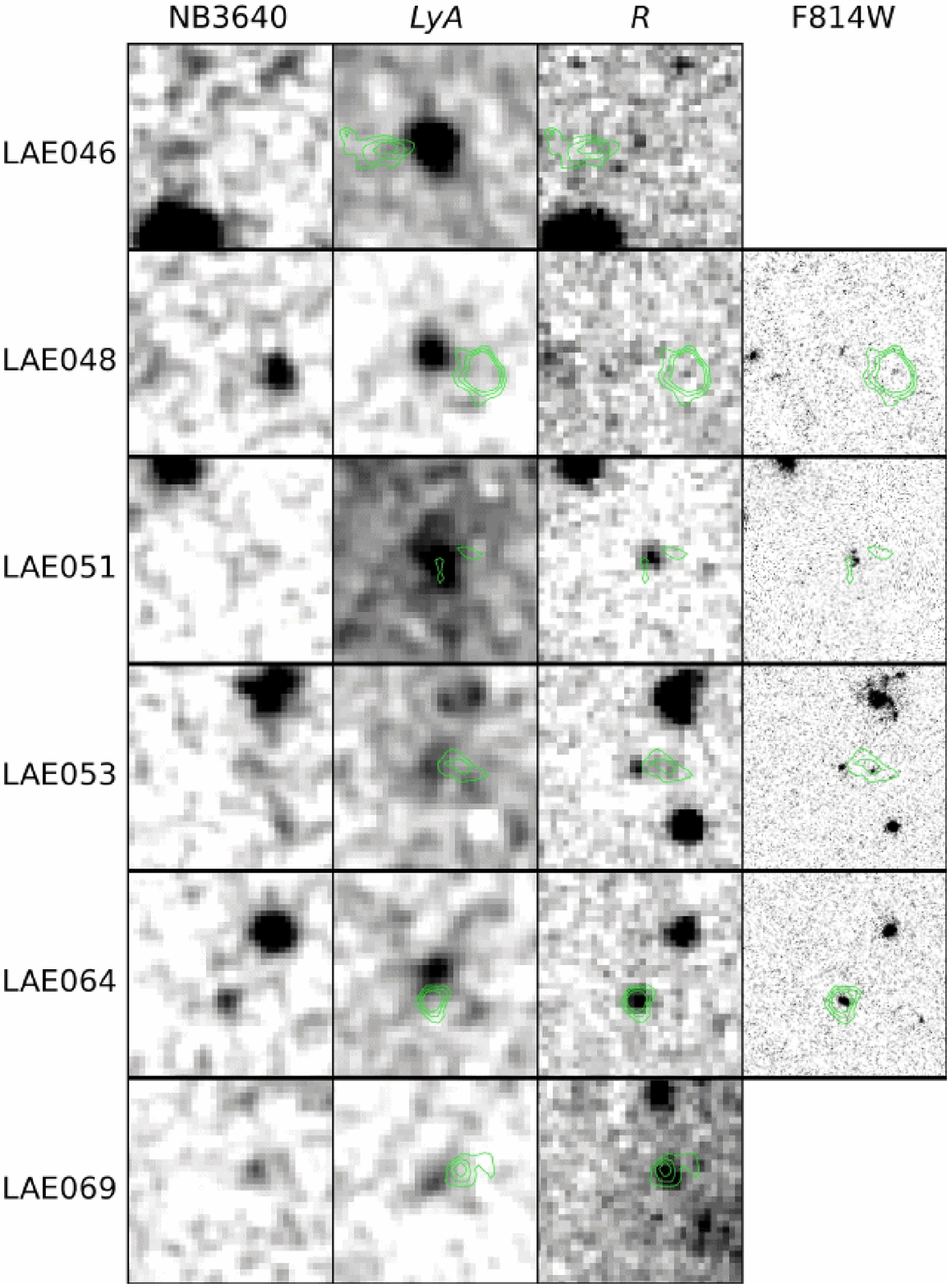}
\caption{\small 
Same as Figure~\ref{f:laes1}. \label{f:laes3}}
\end{figure*}

\begin{figure*}
\epsscale{.826}
\plotone{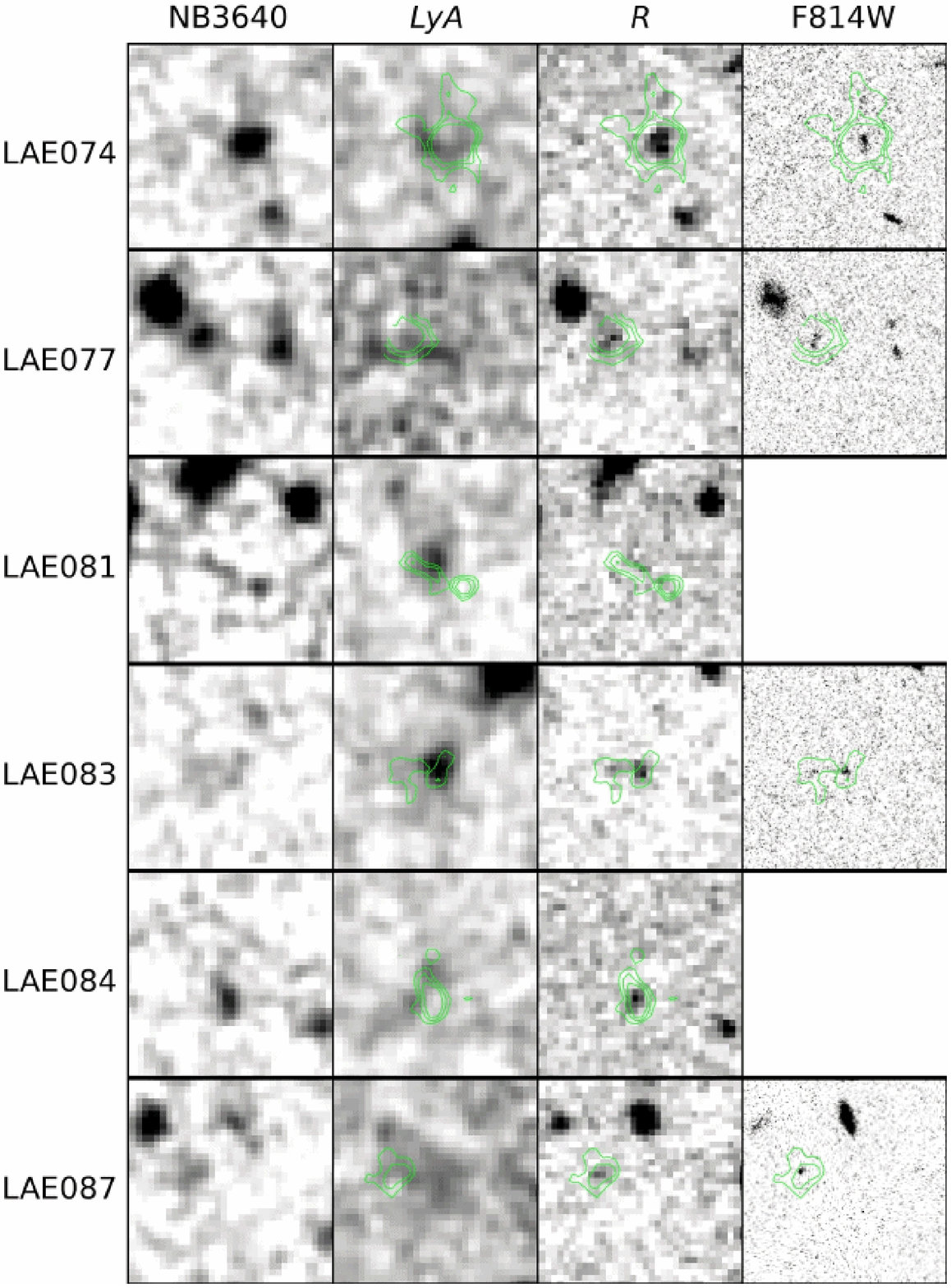}
\caption{\small 
Same as Figure~\ref{f:laes1}. \label{f:laes4}}
\end{figure*}

\begin{figure*}
\epsscale{.826}
\plotone{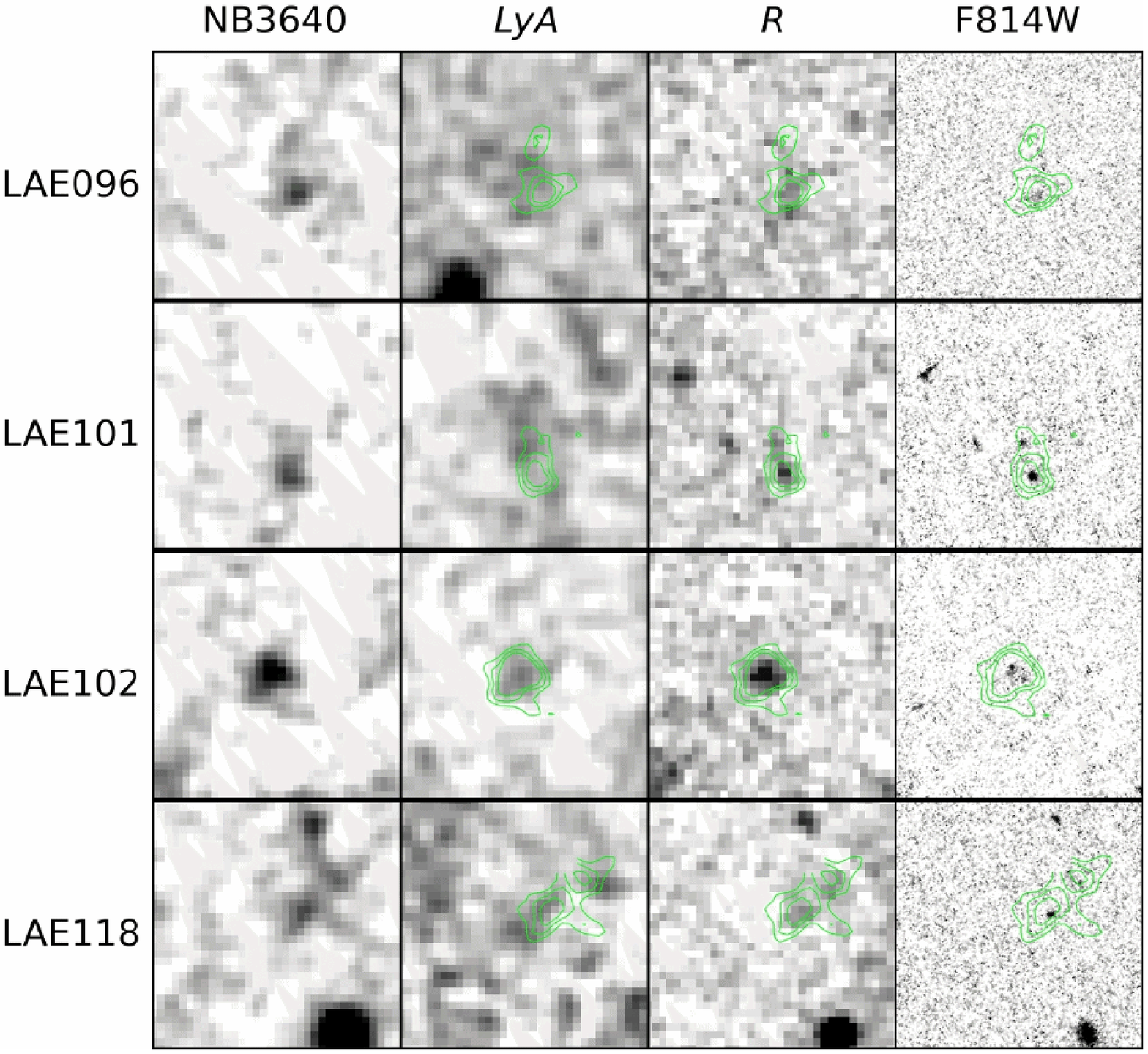}
\caption{\small 
Same as Figure~\ref{f:laes1}. \label{f:laes5}}
\end{figure*}

Figure~\ref{f:lbgs} displays the smoothed NB3640,
$R$ and {\it HST}/ACS-F814W images for the 6
LBGs with NB3640 detections.  Figures~\ref{f:laes1} - \ref{f:laes5} display the
smoothed NB3640, $LyA$, $R$ and {\it HST}/ACS-F814W images for
the 28 (main and faint sample) LAEs with NB3640 detections.  The contours in these Figures
represent NB3640 flux levels corresponding to 28.81, 28.06 and
27.62~mag~arcsec$^{-2}$ (1-, 2- and 3-$\sigma$ pixel-to-pixel
fluctuations in the smoothed image).  

 
We also investigated the average $R$-band and NB3640 flux of various subsamples of LBGs and
LAEs by stacking postage stamp images of the individual galaxies.   Regions
containing flux from sources obvious neighbors were removed
using masks created from the SExtractor
``segmentation'' images.  Both the $R$-band and NB3640 postage stamps were centered on the $R$ detection for LBGs
and on the NB4980 detection for LAEs and, after the rejection of
masked pixels, averaged to create each stack.  No significant
NB3640 flux was detected in either the stack of the 20 LBGs (down to a $2\sigma$ limiting
magnitude of 28.14) or of the 83 LAEs (down to a $2\sigma$ limiting
magnitude of 28.91) lacking individual NB3640
detections.\footnote{Although we did not test the NB3640 filter in the
  lab for the possibility of a ``red leak'', the strong detection
  limits in the stacks indicate that there is no such leak at a level
  that could affect our results.}  The
results from the photometry on stacked subsamples are reported in
Table~\ref{t:stack}.

\begin{deluxetable*}{lccccccccc}
\tablewidth{0pt} 
\tabletypesize{\scriptsize}
\tablecaption{Photometry in stacked-images. \label{t:stack}}
\tablehead{
 \colhead{sample} & \colhead{num.} & \colhead{$R$} & \colhead{NB3640 \tablenotemark{a}} & \colhead{$\sigma_{\mathrm{NB3640-R}}^{\mathrm{phot}}$ \tablenotemark{b}} & \colhead{$\sigma_{\mathrm{NB3640}-R}^{\mathrm{sample}}$ \tablenotemark{c}} & \colhead{NB3640$-R$} & \colhead{$\left<F_{UV}/F_{LyC}\right>_{obs}$ \tablenotemark{d}}
}
\startdata
LBGs & 26 & 24.23 & 27.94 & 0.41 & 0.91 & $3.71 \pm 1.00$ & $30.5^{+46.1}_{-18.3}$\\
LBGs, non-detect & 20 & 24.32 & $>28.14$ & \nodata & \nodata & $>3.82$ & $>33.7$ \\
LBGs, detect & 6 & 23.98 & 25.81 & 0.13 & 0.23 & $1.83\pm 0.26$ & $5.4^{+1.5}_{-1.2}$\\
&&&&&\\
 \tableline 
&&&&&\\
LAEs  & 110 & 26.14 & 27.49 & 0.14 & 0.32 & $1.35 \pm 0.35$ & $3.5^{+1.3}_{-1.0}$ \\
LAEs, non-detect & 83 & 26.19 & $>28.91$ & \nodata & \nodata & $>2.72$ & $>12.2$ \\
LAEs, detect & 27 & 26.04 & 26.22 & 0.14 & 0.57 & $0.18\pm0.59$ & $1.2^{+1.9}_{-0.5}$ \\
LAEs, $R< 25$ & 8 & 24.44 & 27.16 & 0.36 & 0.94 & $2.72 \pm 1.01$ & $12.2^{+18.8}_{-7.4}$ \\
LAEs, $25<R< 26$ & 23 & 25.64 & 26.69 & 0.15 & 0.26 & $1.05 \pm 0.30$ & $2.6^{+0.8}_{-0.6}$ \\
LAEs, $26<R< 27$ & 43 & 26.68 & 27.59 & 0.27 & 0.53 & $0.91 \pm 0.59$ & $2.3^{+1.7}_{-1.0}$ \\
LAEs, $R> 27$ & 36 & 27.53 & 28.41 & 0.67 & 0.84 & $0.88 \pm 1.07$ & $2.2^{+3.8}_{-1.4}$\\
&&&&&\\
 \tableline 
&&&&&\\
 faint LAEs & 20 & 27.93 & $>28.13$ & \nodata & \nodata & $>0.20$ & $>1.2$\\
\enddata
\tablenotetext{a}{\mbox{Limits,} when quoted, are 2$\sigma$.}
\tablenotetext{b}{Photometric color uncertainty determined from stacks of randomly sampled blank regions of the $R$ and NB3640 images.}
\tablenotetext{c}{Sample color uncertainty determined by bootstrap resampling of the subsample.}
\tablenotetext{d}{Observed ratio and uncertainty in non-ionizing UV and LyC emission, inferred from the NB3640$-R$ color of subsample stacks.  This value has not been corrected for either contamination by foreground sources or IGM absorption.}
\end{deluxetable*}

\begin{figure*}
\epsscale{1.10}
\plottwo{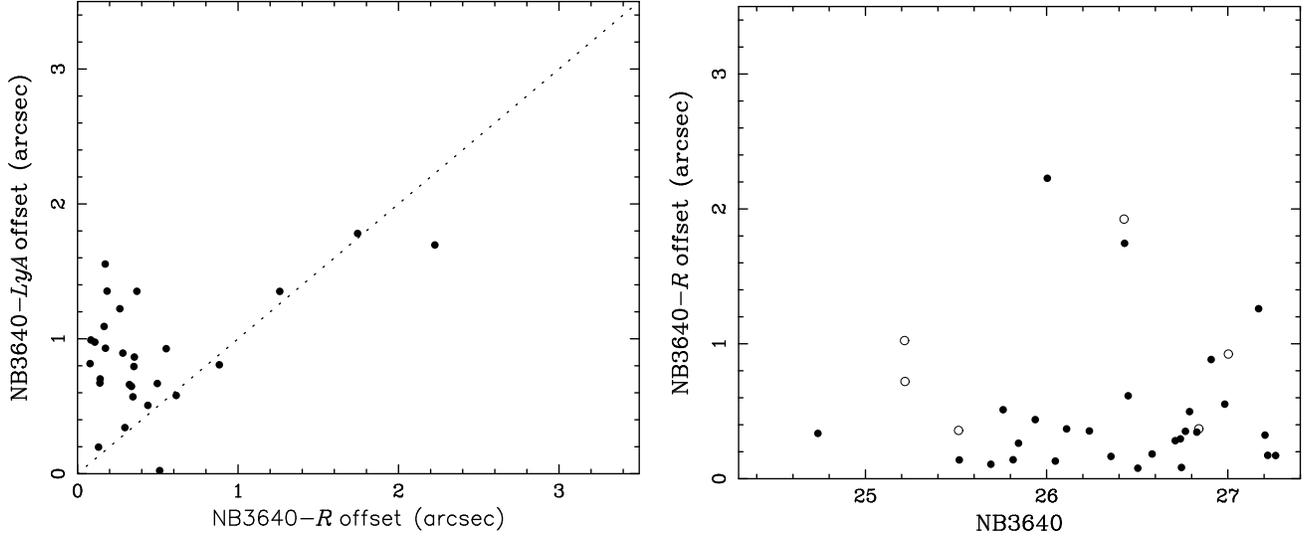}{fig9b.eps}
\caption{\small 
Left: Comparison of the spatial offset between detections in
  the NB3640 and $LyA$ images with those between NB3640 and
  $R$ for our LAE sample.  The smaller NB3640-$R$ offsets relative to
  NB3640-$LyA$ strongly favors the sources being LAE emitters at
  $z\simeq3.09$ over, e.g., [OII] emitters at lower redshift (see text).
  Right: NB3640-$R$ spatial offsets versus NB3640 magnitude.
  Open points represent LBGs while solid points represent LAEs.\label{f:disp1}}
\end{figure*}

As can be seen in Figures~\ref{f:lbgs} - \ref{f:laes5} 
(as well as Tables~\ref{t:lbgs} and \ref{t:laes}), the
NB3640, Ly$\alpha$, and $R$-band fluxes are not precisely
co-spatial in many of our sources.  Spatial offsets between
detected ionizing and non-ionizing UV continuum fluxes
(i.e., significant values of $\Delta_R$), have been previous noted
\citep{iwata2009,inoue2010b}.  The values of $\Delta_R$ in our LAE
sample are relatively small ($\lesssim 0\secpoint6$), however, compared to 
the values of $\Delta_{LyA}$ and $LyA$-$R$ offsets
($\Delta_{LyA,R}$).  The median values for main sample LAEs with
NB3640 detections are $\Delta_R = 0\secpoint32$, $\Delta_{LyA} =
0\secpoint82$ and $\Delta_{LyA,R} = 0\secpoint58$.  We show the
individual values of $\Delta_{LyA}$ versus $\Delta_R$, as well
as $\Delta_R$ versus NB3640 magnitude, in Figure~\ref{f:disp1}. 

\subsection{The source of the NB3640 detections}
\label{s:contam}
The detections in the NB3640 image coincident with the locations of $z\ga 3.06$ LBGs and LAEs
are likely due to some combination of: (i) photons of wavelengths shortward of the
Lyman limit escaping from the
galaxies and being redshifted above the Lyman break before the neutral
phase of the IGM is able to absorb them (ii)
non-ionizing flux from lower-redshift galaxies randomly coincident on the sky
with the high-$z$ systems.  In principle, the high spatial resolution
of the {\it HST}/ACS imaging available for many of our sources with NB3640 detections could help
distinguish between the scenarios.  However, galaxies
observed in the rest-frame UV wavelengths often appear  
clumpy \citep{lotz2006,law2007,peter2007} -- i.e., as distinct regions significantly brighter than the regions
surrounding them -- particularly at high redshift.  
Thus, even with the resolution afforded by {\it HST}/ACS imaging, it is not
generally possible to distinguish in individual systems between clumpy emission from a
single redshift ($z\approx3$) and multiple sources at different redshifts.  

\begin{figure*}
\epsscale{0.80}
\plotone {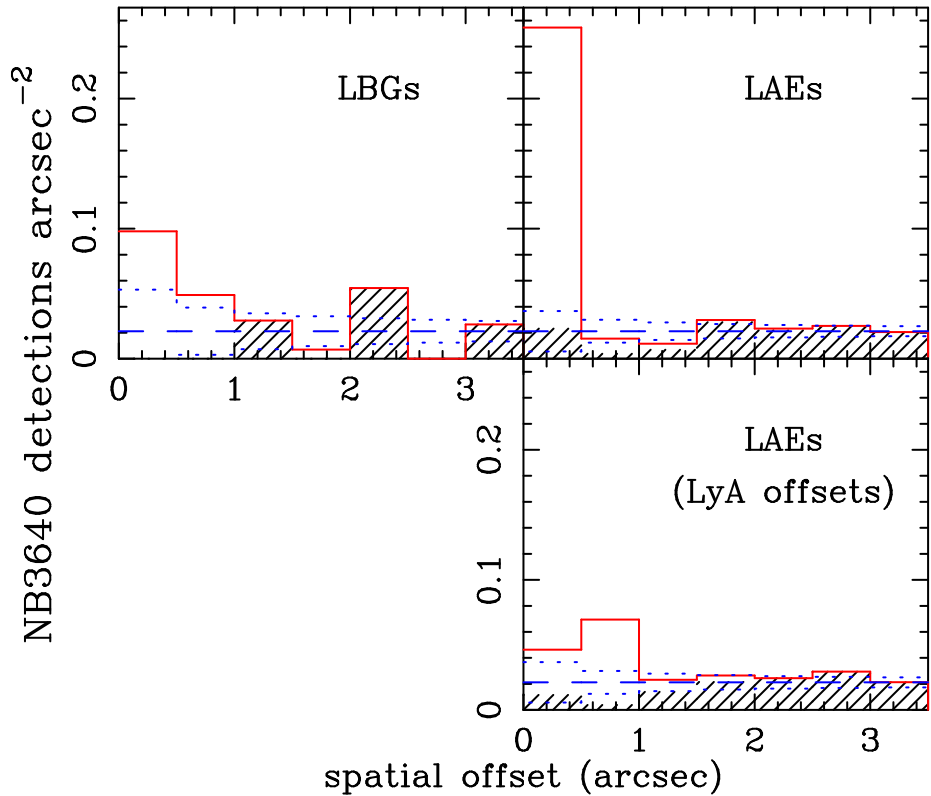}
\caption{\small 
The radial surface-density of NB3640 detections around galaxies in our
LBG and LAE samples.  The solid histograms includes {\it all}
  sources identified by SExtractor; the subset of these sources associated
  with obvious neighbors (which we have
  excluded from our catalog, see \S\ref{s:match}) are
  represented by the hatched region.  The dashed lines indicates the
  global surface density of sources in our NB3640 magnitude range and
  thus represents the expected level of contamination, while the dotted
  lines represent the expected 1$\sigma$ scatter in the contamination.   
  The excess surface density at low offsets indicates that several of
  our low offset LBG NB3640 detections and the bulk
  of the low offset LAE NB3640
  detections are physically associated with the $z\simeq 3.09$ sources
  and not random foreground interlopers.  The top panels use the displacement
  of the LyC centroid from the that of the $R$-band detection (i.e., $\Delta_R$)  for LBGs
  and LAEs with $R$-band detections, the displacements from the
  $BV$ centroid ($\Delta_{BV}$) for LAEs detected in $BV$ but not $R$, and from the
  NB4980 ($\Delta_{\mathrm{NB4980}}$) detection for the remaining LAEs.
  The bottom panel uses displacements from the $LyA$ detections ($\Delta_{LyA}$) of LAEs.
  The excess surface density is contained within a larger range of offsets in
  this panel, as the $\Delta_R$ and $\Delta_{BV}$ values tend to be smaller than the
  $\Delta_{LyA}$ values in our LAE sample (see Figure~\ref{f:disp1}).
\label{f:disp2}}
\end{figure*}

It is possible to address the issue of foreground contamination
in a statistical manner, however.  The positions of any
foreground objects will be uncorrelated with that of the
$z\ga3.06$ sources, and their surface density can be predicted
using, e.g., $U$-band number counts in the ultradeep VLT/VIMOS
catalogue \citep{nonino2009}.  
We predict the surface density of such sources, $\rho_S$,
from the \citet{nonino2009} number counts by interpolating the values
in Table~1 of \citet{vanzella2010b} for our magnitude range $24.5 < U
<27.25$.  To define the positions of the 26 LBGs, we use the centroids of
the $R$-band detections, which have
confirmed spectroscopic redshifts $z>3.06$.  For the 110 LAEs, we
first use the
centroids of the detections in the $LyA$ image (although, see below),
which are very likely to be in the redshift 
range $3.054 \la z \la 3.120$.  In Figure~\ref{f:disp2} we show the
radial surface-density of {\it all} NB3640 detections surrounding the
$z\ga3.06$ sources, in bins of radial offset (i.e., $\Delta_R$ for the LBGs
and $\Delta_{LyA}$ for the LAEs).  The 
surface density is defined as the number of 
detections divided by $\pi(r_{max}^2 - r_{min}^2) N$ where $r_{min}$ and
$r_{max}$ are the minimum and maximum 
radii of the bin and $N$ is the number of $z\ga3.06$ sources, 26 for
the LBGs and 110 for the LAEs.  The left-hand panel shows the  
distribution for the LBG sample, while the bottom right panel shows
the distribution for the LAEs.  The shaded portion  
of the histograms represent detections that we have
identified as clearly belonging to a neighboring source other than the target
LBG/LAE (and therefore do not appear in Tables~\ref{t:lbgs} and
\ref{t:laes}).  The dashed and dotted lines indicate the predicted
surface density of interlopers, $\rho_S$, and the 1$\sigma$
uncertainty in the prediction, respectively.
At large offset, the distributions are comprised entirely of
obvious neighbors, flat and consistent with
$\rho_S$.  This consistency
indicates that any possible difference 
between the $U$-band number counts and the surface density of NB3640
sources in our field (e.g., from the differences in filters, cosmic
variance, etc.) can be ignored for our purposes.  While
disentangling true detections from interlopers becomes more difficult
at smaller offsets, there is clearly a significant excess of detections above the
average foreground level, indicating that many
of the retained NB3640 detections with offsets $\la 1$\arcsec\ must be
physically related to the redshift $z \ge 3.06$ sources.  
The magnitude of this excess over that predicted by the $U$-band number
counts represents the predicted number of uncontaminated LyC
detections, which we quantify below.

We use $\rho_S$ to compute 
the full (radially-dependent) probability
distribution for the number of interlopers in our samples.
Following \citet{vanzella2010b}, the probability that a given LBG/LAE
has a random foreground source with an offset in the range $r_{min} :
r_{max}$ is then
\begin{equation}
p_r = \pi (r^2_{max} - r_{min}^2) \times \rho_S.
\end{equation}
The probability that $K$ of the $N = 131$ sources have a contaminant
centered within the annulus is 
\begin{equation}
f_r(K) = \left(\begin{array}{c}N\\K\end{array}\right)\,p_r^K\,(1-p_r)^{(N-K)}.
\end{equation}
The expectation value for the number of contaminated LBGs/LAEs is 
\begin{equation}
\left<n_r\right> = \sum_{K} K\, f_r(K).
\end{equation}
Note that, dividing $\left<n_r\right>$ by the areas corresponding to each bin in
Figure~\ref{f:disp2}, one recovers the surface density indicated by the dashed line.

Having determined $f_r(K)$ for bins of offset, we can use Monte
Carlo simulations to make a robust prediction of the number of
uncontaminated NB3640 detections.  In each bin, we randomly select
a number of interlopers, $n_{inter}$, from $f_r(K)$.  We then subtract the
number of previously rejected sources in the bin, $n_{rej}$ (i.e.,
those corresponding to obvious neighbors, and
indicated by the hatched histogram in Figure~\ref{f:disp2}).  If
$N  = n_{inter} - n_{rej} > 0$, we randomly reject $N$ of the retained
detections in the given bin.  We repeated the simulation 1000 times to
determine the average and $1\sigma$ deviation of uncontaminated NB3640
detections.   The results of our simulation
indicate that $3.2 \pm 1.2$ of the 26 LBGs and $18.0 \pm 3.2$
of the 110 main sample
LAEs have NB3640 detections that are uncontaminated by foreground
sources.  The two samples are not independent, as they share 5
sources; considering the sample as a whole, our simulation determines 
that $21.6 \pm 3.8$ of the 131 sources have uncontaminated NB3640 detections.
These same simulations can also be used to contamination-correct 
the ensemble average NB3640 magnitudes determined from the stacked images
(Table~\ref{t:stack}) or by summing the individual magnitudes assuming
zero flux from the non-detections (see \S\ref{s:jnu}).  

As discussed in \S\ref{s:dets}, detections in the NB3640 images associated with 
LAEs are less spatially consistent with the detected Ly$\alpha$ emission
than with the rest-frame UV continuum emission.  Thus, it will be more
suitable to define the LAE positions using the centroids of the $R$-band
detections (i.e., use $\Delta_R$ in place of $\Delta_{LyA}$) for the
purpose of estimating the contamination, 
unless significant numbers of the NB3640
detections that we have associated with $z\simeq 3.09$ LAEs are
interlopers {\it and} the corresponding $R$-band detection is dominated by the
interloper.  However, as we concluded above that the majority of such
detections are {\it not} contaminated, the $\Delta_R$ offsets should
lead to more realistic predictions of the number of 
uncontaminated sources.  The upper-right
panel of Figure~\ref{f:disp2} shows the radial surface density of
detections using $\Delta_R$ offsets for the 78 LAEs for which
an $R$-band centroid was measurable, $\Delta_{BV}$ offsets for 9 LAEs detected in the $BV$ image
but not $R$, and $\Delta_{\mathrm{NB4980}}$ offsets for the remaining
23 LAEs.  Using these offsets, our Monte Carlo simulations predicts $20.8  \pm 2.7$ 
of the 110 main sample LAEs and $24.5 \pm 3.0$ of the 131
combined (LBG+LAE) sample sources are uncontamined. 

Finally, we consider the possibility that the LAEs with NB3640
detections having small NB3640-$R$ offsets are not at $z \simeq 3.09$,
but rather, e.g., [OII] emitters at $z \simeq 0.34$. 
In this scenario, the NB3640 filter would sample the rest-frame UV
flux and NB4980$-BV$ would be a measure [OII] emission, both of which
should closely trace regions of active star-formation, while $R$ would correspond to
the rest-frame optical.  The observed NB3640 flux would then be more spatially correlated
with NB4980$-BV$ compared to $R$.  As can be seen in Figure~\ref{f:disp1},
however, the opposite is the case. 

\section{Results}
One of the primary goals of this work is to estimate the comoving
emissivity of ionizing radiation that escapes from star-forming
galaxies at $z\sim3$.  Ideally, this estimate would entail deep,
rest-frame LyC imaging of
several large, independent fields to determine the $\lambda \simeq 900$\AA\
luminosity function; however, such an experiment would be prohibitively
expensive in terms of observing time.  Here, we have taken advantage of the large number of 
galaxies at $z\ge3.06$ in the SSA22a field, which contains a known protocluster at $z=3.09$,
to measure the emergent UV flux escaping from a large sample of LBGs
and LAEs both redward and blueward of the Lyman limit.   The resulting
ratios of non-ionizing to ionizing flux densities allow us to use the 
previously measured global non-ionizing UV emissivity at $z\sim3$ to
determine the overall contribution to the ionizing UV background at
these redshifts from galaxies represented by those in our
subsamples.

\subsection{Observed UV to Lyman-continuum flux density ratios}
\label{s:fluxr}
\begin{figure*}
\epsscale{0.9}
\plotone{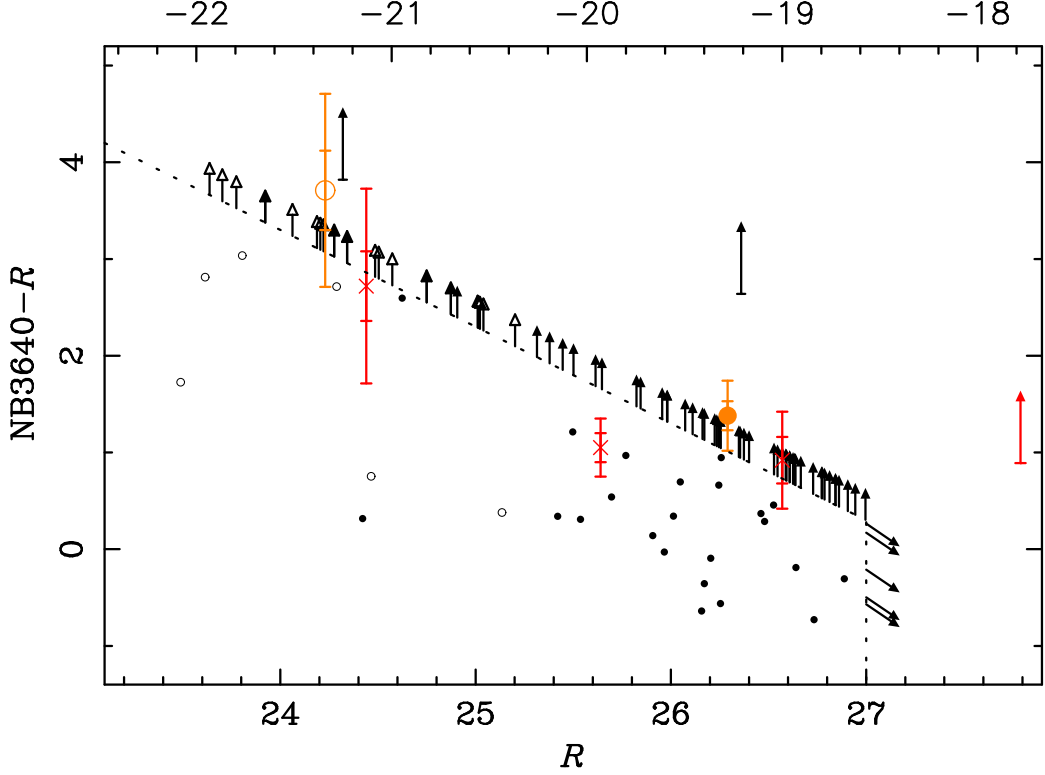}
\caption{\small 
Rest-frame LyC$-$UV color versus rest-frame UV magnitude.  Open
  (filled) points and arrows represent LBGs (LAEs with $R$ and/or NB3640 detections).  
  The uncertainty in color for fainter systems is fairly large; e.g.,
  $\sigma_{\mathrm{NB3640}-R} \ga 0.5$ for $R \simeq 26.5$, NB3640$-R
  \simeq 0$ (see Table~\ref{t:errs}).  Large
 points represent measurements made on stacks of the entire sample of
  LBGs and LAEs.  Lower-limits are for stacks of LBGs
  (far-left arrow) and LAEs (middle arrow) with no individual NB3640
  detection.  Smaller error-bars indicate measurement
uncertainty, while the larger bars also include sample variance
computed through bootstrap resampling.  
Red crosses (and far-right lower-limit) are measurements from stacks in bins
of $R$ magnitude.  Note that five of the eight LAEs in the brightest bin
are also LBGs.  The top axis is the corresponding absolute magnitude scale
for $z=3.09$.\label{f:cmd2}}
\end{figure*}

The Suprime-Cam $R$ filter has an effective wavelength of
$\sim1600$\AA\ and FWHM of $\sim300$\AA\ in the rest-frame of our LBG and LAE samples.
This is is a useful spectral range for determining UV to LyC flux density ratios, as
the luminosity function at $\lambda \simeq 1700$\AA\ has been well
studied at $z\sim3$ \citep[e.g.,][]{reddy2008}.  We show the NB3640$-R$
colors of our sources as a function of $R$ in Figure~\ref{f:cmd2}.  Also
shown are the colors of stacks (\S\ref{s:dets}) of all LBGs, all LAEs,
LBGs and LAEs having no individual NB3640 
detection, and subsamples LAEs in bins of $R$.  Sources with NB3640 
detections in objects brighter than $R \simeq 25$, while few (5 LBGs and
2 LAEs), span a large range NB3640$-R$ color ($\simeq 0$ - 3).  While the
apparent trend in color with $R$ is in part a result of the NB3640 detection
limit (indicated by the dotted line in Figure~\ref{f:cmd2}) and the
increased photometric uncertainty for fainter systems (e.g.,
$\sigma_{\mathrm{NB3640}-R} \ga 0.5$ for $R \simeq 26.5$, NB3640$-R
\simeq 0$; see Table~\ref{t:errs}), photometry on the stacks of LBGs
and LAEs in bins of $R$ indicates a significant difference in the
average NB3640$-R$ color of systems brighter and fainter than $R \simeq
25$.  

The observed UV to LyC flux density ratios can be computed
directly from the values of NB3640 and $R$.  We present these ratios
in Tables~\ref{t:lbgs} and \ref{t:laes}.  
Our value for C49, $F_{UV}/F_{LyC} = 16.6 \pm 6.1$,
is consistent with that derived from the LRIS spectrum by
\citet{shapley2006}, $F_{UV}/F_{LyC} = 12.7 \pm 1.8$.  However, unlike
\citet{shapley2006} we do
not detect D3 in NB3640, consistent with the findings of \citet{iwata2009}.
The average ratios for the ensembles of LBGs and LAEs are
also of interest.  These can be computed either from the photometry on the
stacked images or from the individual $R$ and NB3640 measurements.
Motivated by the non-detection of NB3640 flux in
the stacks of individually non-detected sources, we assume zero flux
for the non-detections in NB3640 when computing the ensemble ratios
from individual measurements.\footnote{For ease we also assume zero
flux in $R$ for LAEs with no $R$-band detection.  This assumption has negligible effect on our results.}  Doing
so, we find values of $\left<F_{UV} / F_{LyC}\right>_{obs}
= 17.8^{+10.4}_{-6.6}$ and $4.2^{+1.2}_{-1.0}$ for the ensembles of LBGs and main
sample LAEs, respectively.  Considering only sources with NB3640 detections, we find 
$\left<F_{UV} / F_{LyC}\right>^{det}_{obs} = 5.2^{+1.8}_{-1.4}$ for
LBGs and $1.3 \pm 0.3$ for LAEs.  In all cases, the errors are
dominated by the 
sample variance which is determined using bootstrap resampling.  The ensemble
results are summarized in Table~\ref{t:fdr1}.  The ratios determined
from the stacked data, which we report in Table~\ref{t:stack}, are
consistent with the above values but with larger uncertainties.  For
the sources with non-detections, we find limits of $\left<F_{UV} /
  F_{LyC}\right>^{non-det}_{obs} > 33.7$ and $>12.2$ for the LBGs and
LAEs respectively.

\begin{deluxetable*}{lcccc}
\tablewidth{0pt} 
\tabletypesize{\scriptsize}
\tablecaption{Average UV to LyC flux density ratios. \label{t:fdr1}}
\tablehead{
\colhead{} & \multicolumn{2}{c}{LBGs} & \multicolumn{2}{c}{LAEs} \\
\colhead{Correction} & \colhead{$\left<\mathrm{NB3640}\right> - \left<R\right>$ \tablenotemark{a}} &
\colhead{$\left<F_{UV}/F_{LyC}\right>$ \tablenotemark{b}} & \colhead{$\left<\mathrm{NB3640}\right> - \left<R\right>$ \tablenotemark{a}} & \colhead{$\left<F_{UV}/F_{LyC}\right>$ \tablenotemark{b}} 
}
\startdata
&&&&\\
\multicolumn{5}{c}{Full Ensembles} \\
&&&&\\
none & $3.13 \pm 0.50$ & $17.8^{+10.4}_{-6.6}$	 & $1.56 \pm 0.28$ & $4.2^{+1.2}_{-1.0}$ \\
contamination  \tablenotemark{c} & $3.68 \pm 0.69$ & $29.6^{+26.3}_{-13.9}$	 & $1.82 \pm 0.36$ & $5.4^{+2.1}_{-1.5}$ \\
IGM + contamination  \tablenotemark{d} & $2.64 \pm 0.70$ & $11.3^{+10.3}_{-5.4}$	 & $0.86 \pm 0.36$ & $2.2^{+0.9}_{-0.6}$ \\
&&&&\\
\hline
&&&&\\
\multicolumn{5}{c}{Sources with NB3640 detections only} \\
&&&&\\
none & $1.78 \pm 0.33$ & $5.2^{+1.8}_{-1.4}$	 & $0.25 \pm 0.25$ & $1.3 \pm 0.3$ \\
contamination  \tablenotemark{c} & $1.65 \pm 0.51$ & $4.6^{+2.7}_{-1.7}$	 & $0.28 \pm 0.30$ & $1.3^{+0.4}_{-0.3}$ \\
IGM + contamination  \tablenotemark{d} & $0.45 \pm 0.59$ & $1.5^{+1.1}_{-0.6}$	 & $-0.69 \pm 0.32$ & $0.5^{+0.2}_{-0.1}$ \\
\enddata
\tablenotetext{a}{Color determined from average NB3640 and $R$-band fluxes.  Uncertainties include individual flux and sample uncertainties.}
\tablenotetext{b}{Ratio and uncertainty in non-ionizing UV and LyC emission inferred from $\left<\mathrm{NB3640}\right> - \left<R\right>$ color.}
\tablenotetext{c}{Color and flux density ratio after statistically correcting sample for foreground contamination of NB3640 fluxes.}
\tablenotetext{d}{Color and flux density ratio after correcting sample for forground contamination and IGM absorption of NB3640 fluxes.}
\end{deluxetable*}

\subsection{IGM absorption of Lyman-continuum flux}
\label{s:igm}
Any LyC flux that escapes from the galaxies in our samples is 
diminished by absorption 
from the neutral phase of the IGM.  In order to model IGM opacity, 
we use Monte Carlo simulations of intergalactic absorption
to generate large (500) samples of random sightlines
to the redshifts of our target galaxies.
Our simulations are a modified version of those presented
in \citet{shapley2006}. 

Absorbers are drawn at
random from column-density and redshift distributions
consistent with recent determinations in the literature
\citep{inoue2008,faucher2008,songaila2010} for both the lower-column-density
Ly$\alpha$ forest, and higher-column-density Lyman limit systems (LLSs) and
Damped Ly$\alpha$ systems (DLAs). Of particular
relevance for estimating the attenuation just below
912 \AA\ are the statistics for LLSs
($N_{\rm{HI}}\geq 10^{17.2}$). For these systems, we assume a power-law
distribution in column-density, proportional to $N_{\rm{HI}}^{-1.3}$
\citep{inoue2008,prochaska2010}, redshift evolution proportional to
$(1+z)^{2.04}$ \citep{songaila2010}, with a total number per
unit redshift at $z\sim3$ of
$dN/dz\approx2$ \citep{songaila2010, stengler1995,steidel1992_pasp}.
Randomly generated absorbers are then applied to the unabsorbed continuum
of each model sightline until the total number of absorbers
is equal to a random Poisson realization of the integral of the
absorber column-density and redshift distributions over the relevant
ranges in column-density ($N_{\rm{HI}}=10^{12}-10^{22}\mbox{ cm}^{-2}$)
and redshift ($z=1.7-z_{source}$, where $z_{source}$ is the redshift of
the target galaxy). The simulations predict a mean free path
to ionizing radiation at $z\sim 3$ (as estimated from the redshift
at which $F_{cont}/F_{912} = \exp(-1)$, following \citet{prochaska2009}),
which is consistent with recent determinations
in the literature of $\lambda_{mfp}\sim 72-85$ proper Mpc
\citep{faucher2008,prochaska2009,songaila2010}.  Using the relation
given by \citet{songaila2010}, we adopt a value of $\lambda_{mfp} =
75.6$~Mpc at $z=3.1$ for the analysis in \S\ref{s:jnu}.

For each target redshift, the resulting sample of 500 model sightlines is
analyzed to obtain the median and standard deviation of the attenuation
factor within the NB3640 filter. This quantity differs from the
one presented in \citet{shapley2006}, which corresponds to the
attenuation factor in the fixed rest-frame bandpass
of $880-910$~\AA. In \citet{shapley2006},
a uniform transmission function was also
assumed across this fixed rest-frame bandpass.
Here we estimate the attenuation in
a fixed {\it observed frame} bandpass, and also take into
account the shape of the NB3640 filter. For higher redshifts,
the attenuation within the NB3640 filter is stronger not
only due to the  the evolving
physical properties of the Ly$\alpha$ forest,
but also the fact that the NB3640 filter probes
bluer rest-frame wavelengths sensitive to photons that
have traveled a longer path-length from the emitting
galaxy.  

\begin{figure}
\epsscale{1.15}
\plotone{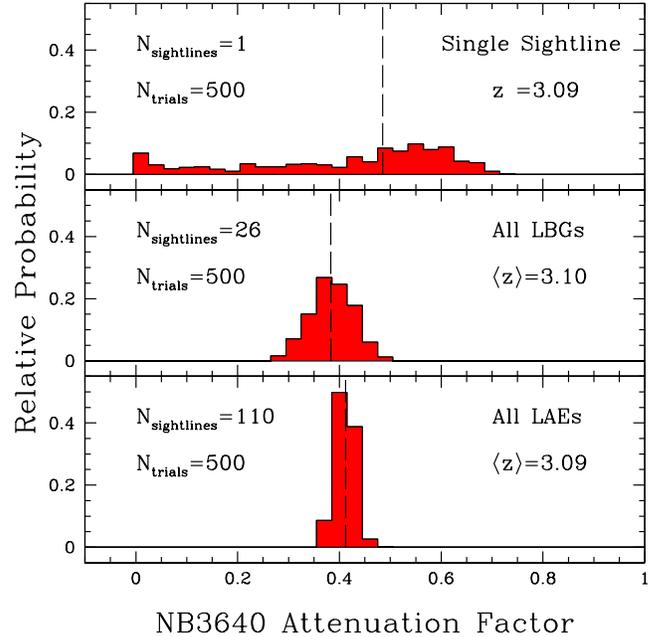}
\caption{\small Results of Monte Carlo simulations of IGM opacity along the line of sight
to our high-redshift targets. To account for IGM opacity, we generated a
large number of random realizations of our measurements through the NB3640
filter. These simulations were tuned to the observation of an individual
sightline (relevant for observations of individual LBGs and LAEs) as well
as the average of ensembles of sightlines (relevant for sample averages
and stacked images). In each simulation, using 500 random realizations, we
recovered the distribution of the ratio between observed and unattenuated
NB3640 flux, denoted here as ``NB3640 Attenuation Factor." In each panel,
the normalized distribution of simulated NB3640 attenuation factors is
plotted, relevant for a given observed quantity. The dashed vertical line
indicates the median NB3640 attenuation factor from each simulation, used
to correct the observed individual or ensemble average $F_{UV}/F_{Lyc}$
flux-density ratio for IGM attenuation. (Top) The results for a single
sightline to a source at $z=3.09$. (Middle) The results for the ensemble
average of 26 LBGs, with a mean redshift of $\langle z \rangle = 3.10$.
(Bottom) The results for the ensemble average of 110 LAEs, with a mean
redshift of $\langle z \rangle = 3.09$.
\label{f:igm}}
\end{figure}

For the six LBGs with NB3640 detections, the median
and standard deviation of the 500 attenuation factors determined for
the appropriate redshift are used to correct the observed UV to LyC
flux density ratios.  We assume $z=3.09$ for each LAE.   We use the average
redshift of the LBG sample, $\left<z\right>=3.10$, and
of the subsample with NB3640 detections, $\left<z\right>=3.13$, to
correct the subsample ratios.   In Figure~\ref{f:igm} we show the
distribution of attenuation factors for our $z=3.09$ 
simulation as well as the distributions for the LBG and LAE ensemble
average attenuation factors.
When computing the uncertainties in the median
attenuations for LBG and LAE subsamples, we assume the sightlines
to the individual sources are uncorrelated.  The corrected flux density ratios
are presented in Tables~\ref{t:lbgs} and \ref{t:laes} for individual
sources, and in Table~\ref{t:fdr1} for ensembles.

\subsection{The escape fraction of ionizing photons}
\label{s:fesc}

The UV to LyC flux density ratios determined in our samples of LBGs and LAEs
can be used to estimate the fraction of LyC photons that escape from
galaxies into the IGM, $f^{LyC}_{esc}$.  However, we must first
correct the ensemble average flux density ratios for contamination
(\S\ref{s:contam}) and IGM absorption (\S\ref{s:igm}).  
In each realization of the Monte Carlo simulations
described in \S\ref{s:contam}, we set the NB3640 fluxes to zero for sources
selected as interlopers by the simulation.  We retain all measured $R$-band
fluxes, which is an appropriate approximation if the interlopers are significantly
bluer in observed NB640$-R$ relative to the $z\sim3$ source, and in any
case is the most conservative approximation.  We then apply the
appropriate IGM correction (\S\ref{s:igm}) to the average flux density ratios 
determined from the contamination-corrected samples.  We present the
raw and corrected colors and flux density ratios for the ensembles and
subsamples with NB3640 detections in Table~\ref{t:fdr1}.
Errors include sample variance computed by first randomly reassigning individual
magnitudes based on the measured magnitude and error, assuming
Gaussian magnitude uncertainties determined from our photometric
simulations (\S\ref{s:phot}), and then bootstrap resampling the data
set.  After applying both corrections, we find $\left<F_{UV}/F_{LyC}\right>^{LBG}_{corr}
= 11.3^{+10.3}_{-5.4}$ for our sample of LBGs.  Our sample of LAEs has on
average much bluer LyC$-$UV colors, with a corrected flux density ratio of
$\left<F_{UV}/F_{LyC}\right>^{LAE}_{corr} = 2.2^{+0.9}_{-0.6}$.  
The corrected ratios represent the average relative flux density in the
vicinity of $z\simeq3$ galaxies.  

The absolute LyC escape fraction can
be inferred from the relative escape fraction as $f_{esc}^{LyC} =
f_{esc}^{rel} \times f_{esc}^{UV}$, where $f_{esc}^{UV}$ is the escape
fraction of non-ionizing UV photons and $f_{esc}^{rel}$ is determined
from the corrected flux density ratio:
\begin{equation}
f^{rel}_{esc} 
=
\frac{\left(L_{UV}/L_{LyC}\right)_{intr}}{\left(F_{UV}/F_{LyC}\right)_{corr}},
\end{equation}
and thus
\begin{equation}
f_{esc}^{LyC} = \left(F_{UV}/F_{LyC}\right)_{corr}^{-1} \times 
  \left(L_{UV}/L_{LyC}\right)_{intr} \times f_{esc}^{UV}
\end{equation}
where $\left(L_{UV}/L_{LyC}\right)_{intr}$ is the (unknown) intrinsic ratio
of UV to LyC luminosity densities produced in star-forming regions.
\citet{reddy2008} report an average UV escape fraction of
$f^{UV}_{esc} \sim$ 20 - 25\% in $z\sim3$ LBGs.  While direct
measurements of $\left(L_{UV}/L_{LyC}\right)_{intr}$ are lacking, spectral 
synthesis models predict values of $\sim6$ for
reasonable assumption of stellar population ages, metallicities and
initial mass functions \citep{siana2007}.  We adopt these numbers as our fiducial
values, resulting in expressions for the relative and absolute escape fractions:
$f_{esc}^{rel}  = 6.0 \times
\left(F_{UV}/F_{LyC}\right)_{corr}^{-1}$ and
$f_{esc}^{LyC} = 0.225 \times f_{esc}^{rel}$.  We caution, however, that
sources with escaping LyC flux may also have larger UV escape fractions.  In any case,
both $f^{UV}_{esc}$ and $\left(L_{UV}/L_{LyC}\right)_{intr}$ 
are highly uncertain and likely vary from object to object.  Also note
that our fiducial value for $\left(L_{UV}/L_{LyC}\right)_{intr}$
differs from that used by \citet{steidel2001} and \citet{shapley2006},
who adopt a value of three.

Most of the 20 LBGs with non-detections in NB3640 have only weak
individual upper limits on the relative escape fraction, ranging from $f^{rel}_{esc}
\la 0.5$ to $\sim$unity.  The ($2\sigma$) upper limit on their average
escape fractions are $\left<f^{rel}_{esc}\right>^{non-det} < 0.5$ and
$\left<f_{esc}^{LyC}\right>^{non-det} < 0.10$.  Of the six LBGs with NB3640
detections, three have corrected UV to LyC flux density ratios consistent with 
$f^{rel}_{esc} \simeq 1$.  However, the other three and the
subsample average have NB3640$-R$ colors
suggesting values greater than unity and therefore in conflict with current models.  
If we assume
that LBGs with NB3640 detections preferentially sample clear
sightlines through the IGM and we can thus ignore the IGM correction 
(but retain the statistical contamination correction) we find
$\left<f^{rel}_{esc}\right>^{det} > 0.82$ (1$\sigma)$ and 
$\left<f^{LyC}_{esc}\right>^{det} > 0.18$ for the subsample of LBGs
with NB3640 detections.  For the LBG ensemble (detections and
non-detections) average, the value of $\left<F_{UV}/F_{LyC}\right>^{LBG}_{corr}
= 11.3^{+10.3}_{-5.4}$ obtained by applying both IGM absorption and
contamination corrections implies average  
escape fractions of $\left<f^{rel}_{esc}\right> = 0.53^{+0.47}_{-0.25}$ and  
$\left<f^{LyC}_{esc}\right> = 0.12^{+0.11}_{-0.06}$.  Note that
all of the above uncertainties do not include contributions from the
uncertainties in $\left(L_{UV}/L_{LyC}\right)_{intr}$ or $f_{esc}^{UV}$.

Due to the overall faintness of the LAE sample, we
can place a meaningful lower limit on $f^{rel}_{esc}$ for only a few
of the individual systems having no NB3640 detection.  
For the non-detection subsample average, we 
determine a weak upper limit of $\left<f^{rel}_{esc}\right> \la 1.2$
for our fiducial intrinsic Lyman break of six.
The LAEs with detections have extremely blue  
NB3640$- R$ colors, however, implying individual $f^{rel}_{esc}$
values greater than unity at high significance for all but one system, even if we neglect
corrections for IGM absorption.  These extremely blue colors lead to
an inferred ensemble relative escape fraction that is greater than unity at
$\sim4\sigma$.  However, an intrinsic value of 
$\left(L_{UV}/L_{LyC}\right)_{intr} = 6$ may be high for the LAEs at $z\sim3$, as their
UV luminosity appears to be dominated by particularly young \citep{gawiser2007}
and low-metallicity \citep{ono2010} populations.  Thus, we also consider a
lower intrinsic flux density ratio of three, which leads to $\left<f^{rel}_{esc}\right>
> 0.75$ (2$\sigma)$, and $\left<f^{LyC}_{esc}\right> > 0.17$ for
the ensemble.  We note, however, that as LAEs have been found to exhibit little
reddening \citep{gawiser2007}, $f_{esc}^{UV}$ is likely larger in our LAE sample than
the fiducial LBG value of 20 - 25\%;
thus our determinations of $f^{LyC}_{esc}$ for LAEs should be
considered conservative lower-limits.
Nonetheless, the NB3640$-R$ colors of LAEs with NB3640 detections are
uncomfortably blue, even when neglecting the IGM-absorption corrections.
The reasons for such
small observed flux density ratios \citep[see
also][]{iwata2009,inoue2010a,inoue2010b} in our LAE samples are not
clear.   

\subsection{The space density of ionizing photons at $z\simeq 3.09$}
\label{s:jnu}
Based on the average corrected UV to LyC flux density ratios for
ensembles of LBGs and LAEs (\S\ref{s:fesc}), we can 
compute the
comoving density of ionizing photons that escape from galaxies 
$z\simeq3$, 
$\epsilon_{LyC} = \left<F_{UV}/F_{LyC}\right>_{corr}^{-1}
\epsilon_{UV}$, where $\epsilon_{UV}$ is the comoving luminosity
density in the non-ionizing UV continuum.  The value of $\epsilon_{UV}$ 
can be obtained by integrating the rest-frame UV luminosity function.
Thus,
\begin{equation}
\epsilon_{LyC} = \left<F_{UV}/F_{LyC}\right>_{corr}^{-1} \phi^*\,L^* \,\Gamma(2+\alpha, L_{min}/L^*)
\label{eq:eps}
\end{equation}
where $\Gamma$ is the incomplete gamma function, $L^*$ is the luminosity
corresponding to the characteristic magnitude $M^*$, and $L_{min}$ is the
minimum luminosity to which the LF is integrated.  The $\lambda\sim1700$\AA\ LF has been
determined by \citet{reddy2008} for LBGs at $z \simeq 3$ with Schechter 
function parameters $\phi^*=1.66 \times 10^{-3}$~Mpc$^{-3}$,
$M^{*}_{AB} = -20.84$, $\alpha = -1.57$ and an effective redshift
$z_{\mathrm{\it eff}} = 3.05$.  

The average emergent UV to LyC flux density ratio for our sample of
LBGs was determined from galaxies brighter than the photometric limit for
spectroscopic confirmation of LBGs in the SSA22a field, $R = 25.5$.
This limit corresponds to $\simeq 0.5L^*$ at $z \simeq 3$.  Thus we
adopt $\left<F_{UV}/F_{LyC}\right>^{LBG}_{corr} =
11.3^{+10.3}_{-5.4}$ for galaxies with $\lambda\sim1700$\AA\ continuum
magnitudes $L \ge 0.5L^*$.  The galaxies
in our LAE sample are generally fainter, reaching at least $R \simeq
27$ ($\simeq 0.1 L^*$).  However, LAEs comprise only a subset of faint
$z\sim3$ galaxies and may provide a biased view of the average far-UV
properties of galaxies with $\lambda\sim1700$\AA\ continuum magnitudes
of $0.1L^* \le L < 0.5 L^*$.  We therefore estimate the contribution
to $\epsilon_{LyC}$ from fainter ($0.1L^* \le L < 0.5 L^*$) galaxies using in turn both the flux
density ratio determined from our LBG sample ($\left<F_{UV}/F_{LyC}\right>^{LBG}_{corr} =
11.3^{+10.3}_{-5.4}$) and that determined from
our LAE sample ($\left<F_{UV}/F_{LyC}\right>^{LAE}_{corr} = 2.2^{+0.9}_{-0.6}$).
Using the LBG-determined flux density ratio, we compute 
$\epsilon_{LyC}(L \ge 0.5L^*) = 7.7^{+7.1}_{-3.7} \times
10^{24}$~erg~s$^{-1}$~Hz$^{-1}$~Mpc$^{-3}$ and
$\epsilon_{LyC}(0.1L^* \le L < 0.5L^*) =
9.2^{+13.8}_{-7.2} \times
10^{24}$~erg~s$^{-1}$~Hz$^{-1}$~Mpc$^{-3}$, for a total of
$\epsilon_{LyC}(L\ge0.1L^*) = 16.9^{+15.5}_{-8.1}
\times10^{24}$~erg~s$^{-1}$~Hz$^{-1}$~Mpc$^{-3}$.  Considering instead the ratio
determined from our LAE sample for the fainter galaxies, we 
find $\epsilon_{LyC}(0.1L^* \le L < 0.5L^*) =
47.0^{+17.6}_{-13.7} \times
10^{24}$~erg~s$^{-1}$~Hz$^{-1}$~Mpc$^{-3}$, for a total of 
$\epsilon_{LyC}(L\ge0.1L^*)  = 54.7^{+19.0}_{-14.2} \times10^{24}$~erg~s$^{-1}$~Hz$^{-1}$~Mpc$^{-3}$.

\begin{figure*}
\epsscale{.80}
\plotone{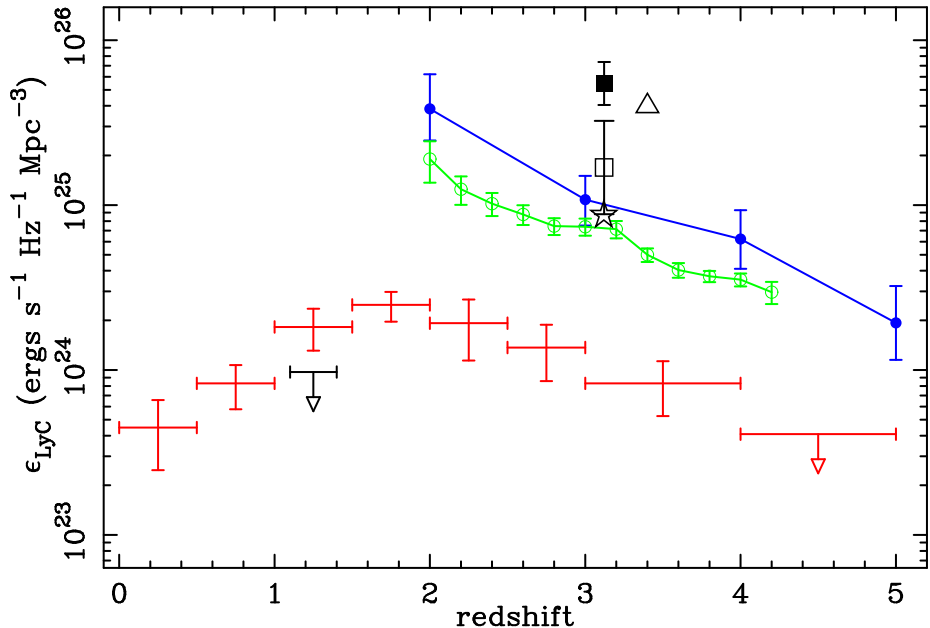}
\caption{\small Contributions to the comoving emissivity of ionizing
  radiation at redshifts $0 < z < 5$.  The open square is our
  determination of the contribution to $\epsilon_{LyC}$ from galaxies
  with $L\ge0.1L^*$ assuming the estimate of the LBG flux density ratio
  holds over this luminosity range, while the filled square adopts our
  estimate of the LAE ratio for galaxies with $0.1L^* \le L < 0.5L^*$.
  Also shown are estimates based on the UV to LyC flux
  density ratios determined from: (star) spectra of 14 LBGs in the
  SSA22a field by \citet{shapley2006}; (triangle) the composite spectrum of 29 LBGs
  by \citet{steidel2001}; and (upper limit) rest-frame $\sim700$\AA\ {\it HST} images of 36
  starburst galaxies at $z\approx1.3$ by \citet{siana2007,siana2010}.
  The red crosses are the
  contribution from broad line QSOs taken from \citet{cowie2009}.
  The open and filled circles are estimates of the total ionizing
  emissivity determined using equations~\ref{eq:jnu} and
  \ref{eq:gamma} and estimates
  of the hydrogen photoionization rate in the Ly$\alpha$ forest, respectively,
  by \citet{faucher2008} and
  \citet{bolton2007}.  Upper limits are 3$\sigma$.
\label{f:eps}}
\end{figure*}

At $z\sim3$, estimates of the UV to LyC flux density ratio in
star-forming galaxies have been made by 
\citet{steidel2001} using a composite spectrum of 29 LBGs and by
\citet{shapley2006} from spectra of 14 LBGs in the SSA22a field. 
We show our estimates of $\epsilon_{LyC}(L\ge 0.1L^*)$ in
Figure~\ref{f:eps}, together with estimates that we derive using the flux
density ratios given by
\citet{steidel2001}  and \citet{shapley2006} and assuming the same
non-ionizing UV luminosity density from $L\ge 0.1L^*$ galaxies that we
computed above using the \citet{reddy2008} LF.  Additionally, using {\it HST}
far-UV imaging of galaxies at $z\sim1.3$,
\citet{siana2007,siana2010} determined (3$\sigma$) lower limits of $\left<F_{UV}/F_{LyC}\right>_{corr} >168$ and 
$\left<F_{UV}/F_{LyC}\right>_{corr} >63$ on the average 
non-ionizing UV to $\sim$700\AA\ flux density ratio for 15
$L\sim L^*$ starburst galaxies in the GOODS fields and 
21 primarily $L<L^*$ 
galaxies in the HDF-N and HUDF, respectively.  We also show the
upper-limit on $\epsilon_{LyC}(L\ge 0.1L^*)$ that we
derive from their flux ratio limits in Figure~\ref{f:eps}, using the
$\lambda \sim 1500$\AA\ LF
parameters determined at $z \sim 1$ from \citet{arnouts2005} and integrating down to $L=0.1L^*$.
The contribution to $\epsilon_{LyC}$ from broad-line QSOs determined
by \citet{cowie2009} is also shown, for comparison.

We can use our estimates of $\epsilon_{LyC}$ to calculate the corresponding
(proper) metagalactic specific intensity, $J_\nu$:
\begin{equation}
J_\nu \simeq (1+z)^3  \epsilon_{LyC} \frac{\lambda_{mfp} }{4\pi}.
\label{eq:jnu}
\end{equation}
We find $J_\nu (L \ge 0.5L^*) = 0.33^{+0.30}_{-0.16} \times
10^{-21}$~erg~s$^{-1}$~Hz$^{-1}$~cm$^{-2}$~sr$^{-1}$ and, assuming the LBG flux
density ratio holds down to $L = 0.1L^{*}$,  $J_\nu (L\ge0.1L^*) =
0.71^{+0.65}_{-0.34} \times
10^{-21}$~erg~s$^{-1}$~Hz$^{-1}$~cm$^{-2}$~sr$^{-1}$. 
Alternatively, computing the contribution from fainter galaxies assuming the flux density
ratio determined from our LAE sample leads to $J_\nu (L\ge0.1L^*)  = 2.30^{+0.80}_{-0.59} \times
10^{-21}$~erg~s$^{-1}$~Hz$^{-1}$~cm$^{-2}$~sr$^{-1}$.  

Furthermore, if we assume a form for the spectrum of escaping
ionizing flux, we can estimate the contribution to the hydrogen photoionization
rate $\Gamma_{\mathrm{H\,{\sc I}}}$ in the IGM from star-forming galaxies at
$z\simeq3$.  The photoionization rate is of particular interest as
$\Gamma_{\mathrm{H\,{\sc I}}}$ has been inferred in an independent
manner from Ly$\alpha$ forest data \citep[e.g.,][]{bolton2005,faucher2008}.  If we assume a power-law spectrum 
$J_\nu = J_{\nu_0} (\nu/\nu_0)^{\alpha_s}$, where
$\nu_0$ corresponds to the Lyman limit and $\sigma_{\mathrm{H\,{\sc
      I}}} = 6.3 \times 10^{-18}$~cm$^{-2}$ is the atomic hydrogen
photoionization cross section at $\nu_0$, then 
\begin{equation}
\Gamma_{\mathrm{H\,{\sc I}}} = \frac{4\,\pi\,\sigma_{\mathrm{H\,{\sc I}}}\,J_\nu}{h\,(3 - \alpha_{s})}
\label{eq:gamma}
\end{equation}
where $h$ is Planck's constant.
The intrinsic spectrum of star-forming galaxies below the Lyman limit
is poorly constrained.  Typical values adopted in past work have
ranged from $\alpha_s = -3$ \citep{bolton2007,ouchi2009} to $\alpha_s
= +0.5$ \citep{faucher2008}.  If we assume a value of $\alpha_s = -3$ we find
a contribution of $\Gamma_{\mathrm{H\,{\sc I}}} = 0.6^{+0.6}_{-0.3} \times
10^{-12}$~s$^{-1}$ from $L\ge 0.5L^*$ galaxies.  For the full range of
$L\ge 0.1L^*$, we determine $\Gamma_{\mathrm{H\,{\sc I}}} = 1.4^{+1.3}_{-0.7} \times
10^{-12}$~s$^{-1}$ using the LBG
flux density ratio, or, computing the contribution for faint galaxies using the
LAE ratio, $\Gamma_{\mathrm{H\,{\sc I}}} = 4.5^{+1.6}_{-1.1} \times
10^{-12}$~s$^{-1}$.  Estimates of $\Gamma_{\mathrm{H\,{\sc I}}}$ from Ly$\alpha$ forest optical depths
range from $\simeq 0.6 \times 10^{-12}$~s$^{-1}$ \citep{faucher2008} to
$\simeq 1.3 \times 10^{-12}$~s$^{-1}$ \citep{kirkman2005}.  The value
obtained by applying the LBG flux density ratio over $L>0.1L^*$ is
consistent with this range, while attributing the LAE flux
density ratio to fainter galaxies results in an estimate that is
$3\sigma$ larger than implied by the
Ly$\alpha$ forest data.  In Figure~\ref{f:eps} we show the values of
$\epsilon_{LyC}$ corresponding to the determinations of
$\Gamma_{\mathrm{H\,{\sc I}}}$ by \citet{bolton2007} and \citet{faucher2008} from z=2
to $z\sim5$.  We arrive at these values using equations~\ref{eq:jnu} and
\ref{eq:gamma} together with our adopted values of $\alpha_S$ and $\lambda_{mfp}$.
Our estimates of $\epsilon_{LyC} $, $J_\nu$, and
$\Gamma_{\mathrm{H\,{\sc I}}}$ are summarized in in Table~\ref{t:jnu}.

\begin{deluxetable*}{ccccc}
\tablewidth{0pt} 
\tabletypesize{\scriptsize}
\tablecaption{Contributions to the ionizing backgrounds. \label{t:jnu}}
\tablehead{
& & \colhead{$\epsilon_{LyC}$ \tablenotemark{c}} & \colhead{$J_\nu$ \tablenotemark{d}} & \colhead{$\Gamma_{\mathrm{H\,{\sc I}}}$ \tablenotemark{e}} \\
\colhead{$\left<F_{UV}/F_{LyC}\right>$ \tablenotemark{a}} & \colhead{Luminosity range \tablenotemark{b}}  & \colhead{($\times 10^{24}$ergs~s$^{-1}$~Hz$^{-1}$~Mpc$^{-3}$)} & \colhead{($\times 10^{-21}$ergs~s$^{-1}$~Hz$^{-1}$~cm$^{-2}$~sr$^{-1}$)} & \colhead{($\times 10^{-12}$s$^{-1}$)} 
}
\startdata 
$11.3^{+10.3}_{-5.4}$ & $L/L^* > 0.5$ & $7.7^{+7.1}_{-3.7}$ & $0.33^{+0.30}_{-0.16}$ & $0.6^{+0.6}_{-0.3}$ \\
$2.2^{+0.9}_{-0.6}$ & $0.1 < L/L^* < 0.5$ & $47.0^{+17.6}_{-13.7}$ & $1.97^{+0.74}_{-0.57}$ & $3.9^{+1.5}_{-1.1}$ \\
\\
\multicolumn{2}{c}{Total} & $54.7^{+19.0}_{-14.2}$ & $2.30^{+0.80}_{-0.59}$ & $4.5^{+1.6}_{-1.1}$ \\
\\
\hline\\
$11.3^{+10.3}_{-5.4}$ & $L/L^* > 0.1$ & $16.9^{+15.5}_{-8.1}$ & $0.71^{+0.65}_{-0.34}$ & $1.4^{+1.3}_{-0.7}$ \\
\enddata
\tablenotetext{a}{Corrected flux density ratio assumed for the luminosity range.}
\tablenotetext{b}{\mbox{Luminosity} range over which the LF is integrated.}
\tablenotetext{c}{Comoving specific emissivity of ionizing radiation.}
\tablenotetext{d}{Proper metagalactic specific intensity of ionizing radiation.}
\tablenotetext{e}{Inferred contribution to the intergalactic hydrogen photoionization rate from galaxies in the luminosity range.}
\end{deluxetable*}

Finally, it should be
noted that the values estimated in this subsection are sensitive to
the adopted parameters.  Specifically, integrating the LF down to $L_{min} = 0$
would result in a factor of two increase in $\epsilon_{LyC}$, $J_\nu$,
and $\Gamma_{\mathrm{H\,{\sc I}}}$ compared to an integration
lower-limit of $L_{min} = 0.1L^*$.  Using the UV LF of
\citet{reddy2009} would not change our estimates 
of the contribution to $\epsilon_{LyC}$ , $J_\nu$, and
$\Gamma_{\mathrm{H\,{\sc I}}}$ from galaxies with $L \ge 0.5
L^*$, but would increase our estimates for the ranges $L \ge 0.1$ and
$0.1L^* \le L < 0.5 L^*$ by 14\% and 27\%, respectively.  Contrastingly,
using the $LyA$ offsets to determine the contamination correction for
the LAEs (\S\ref{s:contam}) would decrease our estimates of the LAE
contribution to $\epsilon_{LyC}$ , $J_\nu$, and
$\Gamma_{\mathrm{H\,{\sc I}}}$ by $\simeq15$\%, while adopting a
far-UV spectral slope of $\alpha_s = +0.5$ would increase our
estimates of $\Gamma_{\mathrm{H\,{\sc I}}}$ by a factor of 2.4.   

\section{Discussion}

\subsection{Summary of the LBG/LAE subsamples}

\subsubsection{LBGs}

Of the 26 LBGs at $z\ge3.06$ in the field of our NB3640 image, we have
detected six down to a limiting magnitude of $27.3$.  Correcting for
foreground contamination and absorption by the neutral IGM, we
determine $\left<F_{UV}/F_{LyC}\right>^{LBG}_{corr} = 11.3^{+10.3}_{-5.4}$.
Previous direct measurements of the relative UV to LyC emergent flux density
ratios at $z\sim3$ have found values ranging from $\simeq 4$
\citep{steidel2001,iwata2009} to $\simeq 20$ \citep{shapley2006}.  Our
value is roughly consistent with both extremes.  For an intrinsic
ratio of six and a non-ionizing UV escape fraction of $\sim 20$ - 25\%, this range
in $\left<F_{UV}/F_{LyC}\right>_{corr} $ corresponds to a range of
0.07 - 0.4 in $\left<f^{LyC}_{esc}\right>$,
with our value giving $\left<f^{LyC}_{esc}\right> = 0.12^{+0.11}_{-0.06}$.   

Monte Carlo simulations based on the distribution of offsets
between the NB3640 and $R$-band detections indicate that two to four of
the LBGs with NB3640 detections are uncontaminated by foreground
sources.  Thus our
corrected LBG detection rate is 8 - 15\% (1$\sigma)$.  Three of the
detections have ratios of non-ionizing to ionizing flux densities that are in
tension with that allowed by stellar population models.  If we take the most
conservative interpretation of our results and consider only C49 and 
D17 -- which have $\Delta_{R} < 1\arcsec$ and $f^{rel}_{esc} \la 1$
-- as uncontaminated detections, we arrive at an ensemble
$\left<f^{rel}_{esc}\right>^{LBG} \simeq 0.10$ and $\left<f^{LyC}_{esc}\right>^{LBG}
\simeq 0.02$.  However, our contamination simulations suggest that
this scenario is overly conservative.    

Other authors have made indirect
measurements of the LyC escape fraction at $z \sim 3$.  \citet{faucher2008} predict
$\left<f^{LyC}_{esc}\right> \sim 5\times 10^{-3}$\ to account for the ionizing background
inferred from the photoionization rate in the Ly$\alpha$ forest.
However, their estimate is averaged over all galaxies down to $L_{UV}
= 0$, assumes an intrinsic UV to LyC flux density ratio of unity, and is
computed using an ionizing spectral slope of $\alpha_s = 
+0.5$; all of these assumptions decrease the escape fraction needed to match
the photoionization rate.  \citet{chen2007} use spectra of long-duration $\gamma$-ray bursts to
determine the column densities of neutral hydrogen along the sightline
though the ISM.  Their results suggest $\left<f^{LyC}_{esc}\right> =
0.02$ in the sub-$L^*$ galaxies that are expected to dominate their
sample.  It is unclear if the difference in the various estimates of
the escape fraction are due to subtleties in the methods or
differences in the samples.  For example, it could be that
$\left<f^{LyC}_{esc}\right>$ depends on environment or galaxy type. 

\begin{figure*}
\epsscale{.85}
\plotone{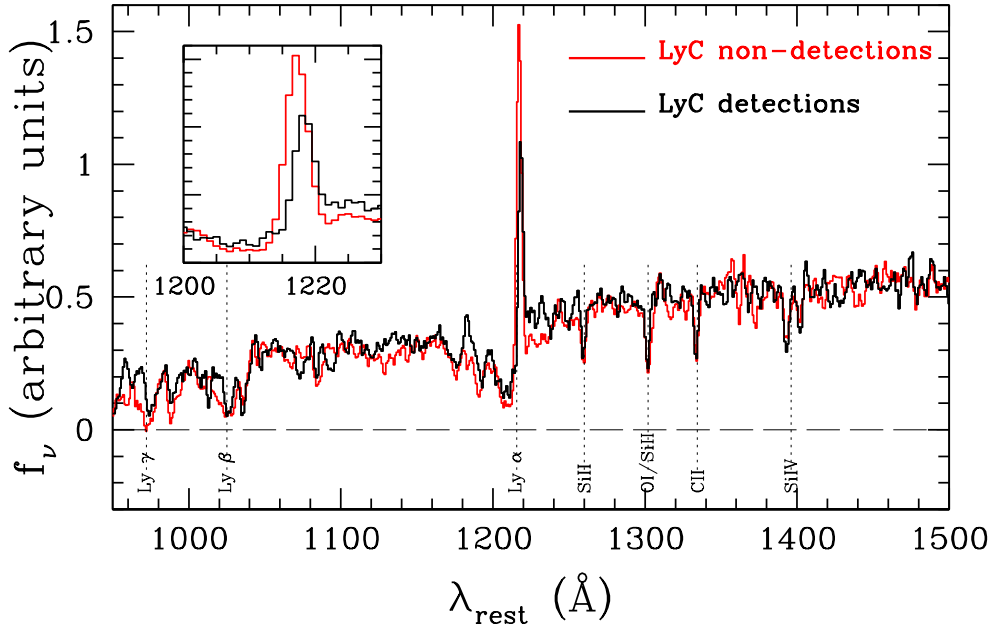}
\caption{\small Comparison of composite rest-frame UV spectra for the six LBGs with LyC
detections in the NB3640 filter (black), and the 20 objects without
detections (red). The composite spectra have been scaled to a common value
over $1400-1500$\AA. In terms of overall spectral shape and strength of
interstellar absorption features, the two spectra are very similar. The
only notable difference is observed in the strength of Ly$\alpha$
emission (indicated in greater detail in the inset panel). The spectrum of
NB3640 non-detections has a Ly$\alpha$ equivalent width $\sim 2-3$ times
larger than that of the detections. At the same time, the centroid of the
Ly$\alpha$ profile in the spectrum of the detections is shifted towards
longer wavelengths.
\label{f:lbgew}}
\end{figure*}

Our sample of NB3640-detected LBGs
is small, and likely $\sim50$\% contaminated by non-detections.  Thus,
any differences in the average rest-frame non-ionizing UV properties between galaxies
with and without leaking LyC flux may not be evident in our data
unless the trends are particularly strong.  Nonetheless, we
investigated the UV properties of the
LBGs in our sample by creating stacks of LRIS spectra \citep{shapley2003,shapley2006}
for galaxies with and without NB3640 detections, which we show in
Figure~\ref{f:lbgew}.\footnote{We note that the offset of the
NB3640 detection from the $R$ band position in some LBGs is significant 
compared to the 1\secpoint2 slits through which the LRIS spectra were
obtained.  Thus, the stellar populations associated with the NB3640
emitting regions may not contribute to the observed spectra in these
cases.}  We find no significant differences between the two
samples in either their average spectral slopes or interstellar absorption lines, although the
strength of the Ly$\alpha$ emission line is on average weaker in our
LBGs with NB3640 detections.  The LBGs undetected in NB3640 are also
slightly fainter on average, consistent with our findings from the
stacked $R$-band photometry (Table~\ref{t:stack}).  We also estimate
the average UV slopes of the two subsamples 
photometrically, both
through aperture photometry on stacked images in the
$V$- and $R$-bands, and through averages of the individual fluxes after
correcting the $V$-band magnitudes for Ly$\alpha$ emission and IGM absorption.  
With neither method do we find significant differences in the average $V-R$
colors between the two subsamples.  We do note that the 26 LBGs in our
sample are on average slightly bluer in $G-R$ (corrected for Ly$\alpha$ emission and IGM absorption)
than LBGs in the large spectroscopic sample presented in
\citet{shapley2003}, which highlights the need for a larger
``average'' sample.

\subsubsection{LAEs}

We detected 27 of the 110 main sample LAEs in the NB3640 image.  Our
Monte Carlo simulations suggest that 18 - 24 of the detections are
uncontaminated,
leading to a corrected LAE detection rate of 16 - 22\%.
After applying contamination- and IGM absorption-corrections we determine
an ensemble $\left<F_{UV}/F_{LyC}\right>^{LAE}_{corr} = 2.2^{+0.9}_{-0.6}$.  Most of our
LAEs with individual detections have flux density ratios inconsistent with
predictions of stellar population models even for young, low-metallicity
populations and neglecting the IGM absorption correction.
In contrast, the lower limit on the average UV to LyC flux density ratio for
LAEs without individual NB3640 detections
($\left<F_{UV}/F_{LyC}\right>^{LAE, non-det}_{obs} > 12.2$) is more
than a factor of 9 higher than the average observed ratio for those
with detections ($\left<F_{UV}/F_{LyC}\right>^{LAE, det}_{obs} =
1.3$), implying a dichotomy in the far-UV properties of $z\simeq3$
LAEs.  

\begin{figure*}
\epsscale{0.85}
\plotone{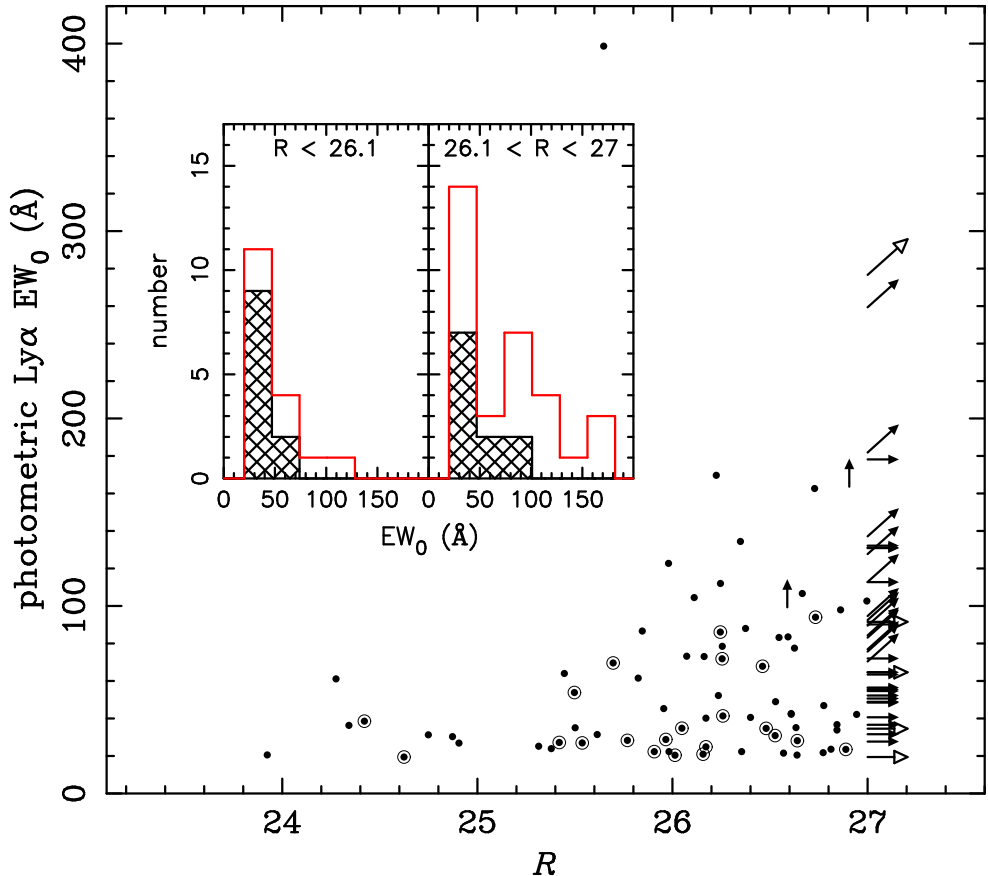}
\caption{\small LAE Ly$\alpha$ rest-frame equivalent widths 
  determined from $BV-$NB4980 colors.  Sources detected in
  NB3640 are indicated by circled points and open arrows (plotted
  slightly larger than solid arrows for clarity).  Consistent with previous
  studies, we find that the highest equivalent widths tend to be
  associated with sources that are
  fainter in the rest-frame UV continuum.  The inset shows the
  distribution of equivalent widths for brighter (left) and fainter
  (right) LAEs; the median $R$-band magnitude for sources with $R$-band
  detections is $R=26.1$.  Sources with and without NB3640 detections are
  represented by the black hashed and red histograms,
  respectively.  
\label{f:laeew}}
\end{figure*}

The implied relative escape fractions $f^{rel}_{esc} \gtrsim 1$ for
our LAEs detected in NB3640 are difficult to explain.  Exotic stellar 
population models with top-heavy IMFs and extremely low metallicities
could in principle reproduce the observed flux density ratios
\citep[e.g.,][]{inoue2010b}, though we caution that 
there is little additional evidence to suggest such models are
appropriate for $z\sim3$ LAEs.  As with the LBGs, we find no
significant differences between the average 
non-ionizing UV slopes of LAEs with and without NB3640 detections,
determined either via aperture photometry on the $R$- and $V$-band
stacks, or via Ly$\alpha$-corrected averages of the individual $V$ and $R$
magnitudes.  In Figure~\ref{f:laeew} we show the Ly$\alpha$ rest-frame
equivalent widths for our LAE sample, estimated from their $BV-$NB3980
colors, as a function of UV continuum (i.e., $R$-band) magnitude.  The
largest equivalent widths tend to be confined to fainter continuum
sources, consistent with past findings \citep[e.g.,][]{ando2006,stark2010}.  In order to be
detected in NB3640, the faintest sources require lower non-ionizing UV
to LyC flux density ratios (e.g., see Figure~\ref{f:cmd2}).  Thus, any
comparison of 
the distributions of equivalent widths between sources with and
without NB3640 detections should be confined to a relatively narrow
range of continuum magnitude.  The inset in Figure~\ref{f:laeew} shows
such distributions for the ranges $R<26.1$ and $26.1 < R < 27$,
demonstrating that
the median Ly$\alpha$ equivalent widths are smaller
for LAEs detected in NB3640 compared to those without
detections.  

\subsubsection{Faint LAEs}

In addition to our main sample LAEs, we have identified 20 
``faint'' LAE candidates with $26<$NB4980$\le26.5$ and
$BV-$NB4980$\ge1.2$.  These faint LAEs have larger photometric
uncertainties (e.g., see Table~\ref{t:errs}).  As they have different
selection criteria from the main sample and are detected at
lower significance, we have not included them in our statistical
analyses.  However, for completeness we present results for the average properties of
this sample from
our stacking analysis in Table~\ref{t:stack} and for individual
objects in Table~\ref{t:laes2}.  We detect only one object in the faint sample in
NB3640, while in the main sample we detect 14\% of the 65 LAEs having
$BV-$NB4980$\ge1.2$ in the NB3640 image.  However, the faint LAEs are also
on average fainter in $R$ and therefore have smaller detection limits in $F_{UV}
/ F_{LyC}$.  Thus, the data do not allow us to address any statistical
differences between the two populations. 

\subsection{Other considerations}

\subsubsection{Are we missing low $\left(F_{UV} / F_{LyC}\right)_{obs}$ LBGs?}

Of the 110 LAEs in our sample, 24\% have NB3640$-R <1.5$, compared
to 8\% of our LBGs.  The Lyman break technique
specifically selects against galaxies with notably blue broadband
$U_n-R$ colors \citep{steidel2003},
and while it is possible for a $z\ge3.06$ galaxy to have
significantly higher flux density through the narrow NB3640 filter
than through the relatively broad $U_n$-band, we note that the effective
wavelengths of the two filters are very similar.
The LAEs discussed here, on the other hand, are selected by their
Ly$\alpha$ emission line strength independently of their 
rest-frame UV colors and corresponding $\left(F_{UV} /
  F_{LyC}\right)$ values.  Therefore we must consider
whether the very application of the Lyman break technique selects
against a significant number of non-LAE galaxies
with $R$ magnitudes similar to our LBG sample but
NB3640$-R$ colors similar to those observed in our LAE sample.
If such a population exists at $z\sim3$, it would appear as a
high-redshift tail in samples of
BX/BM galaxies \citep{steidel2004,reddy2008} selected to target
galaxies with bluer $U_n - R$ colors at $1.4 \la z \la 2.5$.  However,
the fraction and associated surface density of such
galaxies with $z\ge3.06$ is exceedingly small \citep[see, e.g., Figure~1
of][]{reddy2008}, and the incidence of such systems is likely consistent
with foreground contamination.  Thus, while the nature of the high-$z$
BX/BM galaxies should be explored further, it does not appear that 
we are missing significant numbers of rest-frame
UV bright galaxies having NB3640$-R \sim 0$ due to our selection of $z
\sim 3$ galaxies using the Lyman break technique.  
 
\subsubsection{Viewing angle effects}

As with the bulk of our $z\simeq3$ sample, searches for emerging LyC
photons from galaxies at lower redshifts have largely yielded null results
\citep{malkan2003,siana2007,siana2010,bridge2010}.  The small fraction of star-forming
galaxies with detected LyC emission implies that, for ionizing radiation to escape,
there must be pathways through the neutral ISM from star-forming regions that are
unusually clear of dust and gas.  Processes capable of removing or
destroying neutral gas and dust (e.g., feedback, tidal stripping,
etc.) in actively star-forming galaxies are unlikely to be
sufficiently effective to allow the  transmission of ionizing
radiation over all solid angle.  Large spatial variations in $f^{LyC}_{esc}$
are also predicted by simulations
\citep[e.g.,][]{razoumov2007,gnedin2008}.  If LyC flux only escapes over a small
fraction of the total solid angle relative to the non-ionizing flux,
this would naturally explain the large spread in the observed far-UV
properties in our samples.

We consider a simple model in which most of the UV and LyC flux
emitted in directions of cleared ISM escapes while some UV and no
LyC flux escapes in other directions.  If the projected area of the
cleared ISM is comparable to the size of the LyC-emitting 
region, the relative escape fraction {\it from that region} would either be
negligible or near unity, depending on whether the viewing angle
samples a direction through the cleared ISM.  An ensemble
of galaxies having a single, compact starburst would then be expected
to exhibit a bimodal observed distribution
of relative escape fractions.  In contrast, a single projected view of a 
galaxy having multiple or extended regions of star formation could
sample regions both cleared of and obscured by neutral ISM,
effectively averaging out the bimodality inherent in compact systems.   
A wider range of observed values for the relative escape fraction would then 
be possible for such a galaxy.
Based on the ACS-F814W
images in Figures~\ref{f:lbgs} - \ref{f:laes5}, the LAEs in our sample appear to be
on average more compact than the LBGs and, consistent with this simple
picture, more bi-modal in their LyC properties relative to the LBGs. 

If the escape of LyC photons is highly anisotropic, then the
observed UV to LyC flux density ratio in an individual
source is unlikely to be representative of the ratio of the 
integrated (i.e., over all solid angle) UV to
LyC flux densities escaping from that source.  For a sample large enough to
contain a significant number of LyC detections, however, the ensemble average
ratios should reflect the average integrated flux density ratios.  Additionally, under the
assumption that all galaxies in each subsample have similar properties,
the detection rate should indicate the average fraction of solid angle
over which LyC photons can escape a typical star-forming galaxy at $z \simeq 3$.
Of course, if there are evolutionary effects within our LBG and LAE
samples, the assumption of
equivalence within each galaxy sample is not valid.
Careful consideration of the stellar populations of the
sources is required to test this assumption (Steidel et al., in
prep).  For now we note that large differences are not apparent in the
average UV properties of LBGs or LAEs with and without NB3640
detections. 

\subsubsection{IGM coherence effects}

When determining the  uncertainty in the IGM absorption
corrections for the ensemble samples, we assume the IGM properties
along the sightlines to each source are uncorrelated.  
If this is not an adequate assumption, the uncertainties in our
computed escape fractions and IGM-corrected emissivities may be
underestimated.  Indeed, studies have found coherence in the 
Ly$\alpha$ forest over Mpc scales \citep{becker2004,casey2008} at $z\sim 3$,
though correlations are weaker in the higher column
density systems relevant to our corrections and a detailed
description of the sizes and correlation scales of forest clouds is
still lacking \citep[see, e.g.,][]{meiksin2009}.  Furthermore, there
may be systematic environmental effects that we are not considering.  For example,
the fact that our targets lie in or behind a large-scale galaxy overdensity \citep{steidel2000}
may cause the IGM to be on average more opaque than expected with a
higher than average incidence of LLSs.
Alternatively, a proximity effect due to the relatively high density of
star-forming galaxies and QSOs in the protocluster
could in principle diminish the IGM opacity on Mpc scales making the
sightlines to our targets less opaque to ionizing radiation.  It should be
noted, however, that the UV to LyC flux density ratios in our LAEs with
detections are uncomfortably small even before the IGM corrections are
made.  

\subsection{Implications}

Our results suggest the following picture.  Neutral gas and
dust in the ISM of typical star-forming galaxies at
$z\sim 3$, as traced by LBGs, reprocesses most of the
non-ionizing UV flux, and effectively all of the LyC
flux, over most solid angle.  Over some significant fraction of solid
angle ($\sim$10 - 20\%, as suggested by our corrected detection
rates), however, the neutral gas and
dust have been cleared or destroyed, allowing both ionizing and
non-ionizing flux to escape into the IGM in those directions.  Such
regions are likely to be smaller than the projected UV sizes of
typical bright galaxies but may be comparable to the sizes of
LyC-emitting regions in compact sources.  In this picture, {\it when detected in LyC},
compact sources should on average exhibit larger 
relative escape fractions compared to LyC-detected galaxies with extended
star-forming regions.
Sightlines to relatively faint, low 
metallicity star-forming sources, such as may be selected by our LAE
sample, could have lower intrinsic UV to LyC ratios further boosting
the observed non-ionizing to ionizing UV flux density ratios.   

We are left with two 
difficulties in this picture, however.  First, the observed  LyC flux density
relative to the non-ionizing UV continuum for the majority of our
sources with putative LyC detections exceeds predicted
unattenuated ratios for current reasonable stellar population models.
Second, even if such ratios were possible to explain, the implied
contribution to the intergalactic neutral hydrogen photoionization rate
from LAEs is in excess of that determined from measurements of the opacity in
the Ly$\alpha$ forest \citep{bolton2005,faucher2008}.  A satisfactory model that can reproduce all of
the observational evidence is still wanting and follow up spectroscopy \citep[e.g.,][]{inoue2010b}
is needed to determine the nature of the NB3640 detections associated
with our LAEs. 

The LyC photons
that escape from galaxies prior to the end of reionization cannot be
directly observed, as they are absorbed during the reionization
process \citep{fan2006,robertson2010}.  Even after the epoch of reionization, at redshifts $z\ga
4$, the opacity of the IGM to ionizing photons due to overdensities
retaining significant neutral fractions is sufficient to make direct
detection of escaping LyC flux improbable \citep{inoue2008,vanzella2010a}.  Thus,
determinations of $f^{LyC}_{esc}$ from direct measurements at
$z\approx3$ are likely to be the best constraint on the LyC escape
fraction for the galaxies that presumably reionized the Universe.
\citet{gnedin2007} determined that, if the LyC escape fraction for low
mass ($M_{halo} <10^{11} M_{\sun}$) 
galaxies is negligible, there would be insufficient LyC photons for
reionization.  On the contrary our results for LAEs (while
puzzling) imply that relatively faint galaxies on average have
LyC escape fractions larger than those of bright galaxies.
Recent data obtained with the Wide Field Camera 3 on {\it HST}
have allowed for the study of galaxies out to redshifts $z\sim 7$
\citep[e.g.,][]{bunker2010,oesch2010}.  \citet{bouwens2010} find a
steep faint-end slope to the $z\sim7$ LF, $\alpha=-1.94$.  
Parameterizing the observed evolution of the LF from $z\sim4$ to $z\sim
8$, they extrapolate the LF to even higher redshift and claim
that the Universe can be reionized by $z\sim6$ if the average escape
fraction is $\left<f^{LyC}_{esc}\right> = 0.2$.  This result is in agreement
with earlier findings by \citet{ouchi2009}.  If
$\left<f^{LyC}_{esc}\right> = 0.6$, \citet{bouwens2010} claim reionization is
complete by $z\sim8$ and the predicted Thomson
optical depth would be within $1\sigma$ of that measured for the WMAP 7-year
data set.  Since the luminosity density at these redshifts is
apparently dominated by relatively faint galaxies, the LyC escape fraction in 
fainter sources such as our sample of LAEs is of particular
importance.  Resolving the discrepancy between the theoretical minimum
and observed UV to LyC flux density ratios is thus an important undertaking
for future work.

\section{Summary}
We have presented analysis of the deepest near-UV image to date of the
SSA22a  field.  This image was obtained through a narrowband filter
sampling the redshifted Lyman-continuum emission from $z\ge3.06$
sources, including many galaxies in the 
$z=3.09 \pm 0.03$ SSA22a protocluster.  The image contains 26 LBGs with 
spectroscopically-determined redshifts $z\ge3.06$ for which our NB3640
filter is uncontaminated by emission longward of the redshifted Lyman
limit.  We augmented these data with both new and archival
deep narrowband and broadband images designed to select LAEs
at $3.06 \la z \la 3.12$.  From these images we have identified 110
LAE candidates.  Our main conclusions are as follows:

\begin{enumerate}
\item Six of the 26 LBGs have NB3640 detections with AB magnitudes in the
  range 25.2 to 27.0.  Five of the detections are offset from the
  $R$-band detections by $\le 1$\arcsec.  We have used Monte Carlo simulations to determine the
  probability of contamination by foreground objects and conclude that
  two to four (1$\sigma$ confidence interval) of the six detections
  are uncontaminated.  Twenty-seven of the 110 LAEs are detected in NB3640 with
  magnitudes spanning 24.7 to 27.2, twenty of which are offset from
  the $R$-band detection by $\le 1$\arcsec.  Our simulations suggest
  that 18 to 24 of our LAEs with NB3640 detections are uncontaminated.

\item For our sample of LBGs we find a large range in NB3640$-R$
  colors and thus observed UV to LyC flux density ratios.  Furthermore, the
  subsample of LAEs having no individual 
  NB3640 detection has a lower limit on the average UV to LyC flux density ratio
  that is almost an order of magnitude larger than observed for LAEs
  with NB3640 detections,
  implying a significant dichotomy in the observed far-UV properties of LAEs.  We 
  interpret this variety in the observed LyC properties of $z\approx 3$
  galaxies as arising from the partial clearing of
  neutral gas and dust over a limited solid angle in individual
  systems.  In this picture, ensemble measurements of
  $f^{LyC}_{esc}$ (when corrected for foreground contaminants and IGM
  absorption) will produce accurate average values, while measurements
  in individual systems will be highly biased by orientation.

\item We find no difference between the average non-ionizing
  rest-frame UV properties for LBGs with and without NB3640 detections,
  with the possible exception that the Ly$\alpha$ emission line may be
  weaker in those with detections.  We also find no difference in the
  average UV spectral slopes for LAEs with and without NB3640
  detections, although we again find evidence for weaker Ly$\alpha$
  emission in LAEs with detections.  The similar UV properties of the two samples supports
  the idea that orientation effects determine the detectability of
  escaping LyC flux, if such affects are less important for the
  non-ionizing UV compared to the LyC.

\item After applying statistical corrections for foreground
  contaminations and IGM absorption, and assuming an intrinsic
  UV-to-LyC flux density ratio of six and a UV escape fraction of $\sim 20$ -
  25\%, we determine $f^{LyC}_{esc} \sim 0.1$ for  our sample of
  LBGs.  The LBGs in our sample represent the bright end ($L \ga 0.5
  L^*$) of the rest-frame UV LF at $z\approx3$.  If their average UV to LyC
  flux density ratio is representative of LBGs down to $0.1 L^*$, their
  inferred contribution to the intergalactic hydrogen photoionization, $\Gamma_{\mathrm{H\,{\sc I}}}$,
  rate is consistent (within the large uncertainties) with that
  measured from the Ly$\alpha$ forest opacity
  \citep{bolton2005,kirkman2005,faucher2008}. 

\item Half of our LBGs and all but one of our LAEs with NB3640
  detections have IGM-corrected UV to LyC flux density ratios significantly
  smaller than the minimum predicted from stellar population models.
  Our LBGs are drawn from the bright end of the LBG LF, $L \ge 0.5 L^*$.  If the
  average LAE UV to LyC flux density ratio that we measure is 
  representative of LBGs over the magnitude range of our LAEs, $0.1L^*
  \la L \la 0.5L^*$, then the inferred contribution to the global
  hydrogen photoionization rate $\Gamma_{\mathrm{H\,{\sc I}}}$ exceeds
  the values measured from the Ly$\alpha$ forest opacity at
  $z\approx3$.   Further study is needed to determine the nature of
  the LyC properties of these sources. 

\end{enumerate}

Recent measurements of the LBG luminosity function at redshifts $z\ga6$ suggest a large
value of $f^{LyC}_{esc}$ is needed with substantial contribution from
relatively faint galaxies in order for the Universe to be reionized by
$z\sim6$ \citep{ouchi2009,oesch2009,oesch2010,bouwens2010}.  
While the extremely blue NB3640$-R$ colors of our faint LAE
sample are difficult to explain within current theoretical models for
the UV and LyC emission properties of star-forming regions,
reconciling the observations with models is an important step towards
understanding the processes involved in reionization.  To this end, we are
augmenting our data sets with additional {\it HST} imaging  and
ground-based spectra of both LBGs and LAEs with NB3640 detections.
In addition to providing independent confirmation of LyC detections,
such data are needed to confirm the redshifts of the LAEs. 
Finally we note that while our samples of LBGs and LAEs are large
enough to determine robust average properties, it may be that the
SSA22a protocluster is itself unique.  Thus we have ongoing projects
to supplement this work with similar studies in other fields.  

\acknowledgements
We would like to thank Andrew Blain for permitting our use of the 
NB3640 filter, Kevin Hainline for his helpful discussions
regarding our photometric simulations, and the anonymous referee for
constructive suggestions. D.B.N.\ and A.E.S.\ acknowledge
support from the David and Lucile Packard Foundation. C.C.S.\
acknowledges additional support from the John D.\ and Catherine T.\
MacArthur Foundation, the Peter and Patricia Gruber Foundation, and 
NSF grants AST-0606912 and AST-0908805.  We
wish to extend special thanks to those of Hawaiian ancestry on whose
sacred mountain we are privileged to be guests.  Without their
generous hospitality, most of the observations presented herein would
not have been possible. 

\bibliographystyle{apj}
\bibliography{apj-jour,bibliography}

\clearpage

\appendix

\section{LBGs and LAEs with no NB3640 detection}
In this appendix we present coordinates, $R$-band magnitudes and
limits on the observed UV to LyC flux density ratios for the LBGs and LAEs
that were undetected in the NB3640 image, as well as emission and
interstellar absorption redshifts for the LBGs.  Table~\ref{t:lbgs2}
list the LBGs and Table~\ref{t:laes2} lists the LAEs.  The color $BV-$NB4980$=3.49$
corresponds to infinite Ly$\alpha$ equivalent widths; values near this
limit reflect the large photometric uncertainties.

\LongTables

\begin{deluxetable*}{ccccccc}
\tablewidth{0pt} 
\footnotesize
\tablecaption{Photometry for LBGs without NB3640 detections. \label{t:lbgs2}}
\tablehead{
\colhead{ID} & \colhead{RA} & \colhead{Dec} & \colhead{$z_{em}$ \tablenotemark{a}} & \colhead{$z_{abs}$ \tablenotemark{b}} & \colhead{$R$} & \colhead{$\frac{F_{UV}}{F_{LyC}}_{obs}$ \tablenotemark{c}} \\
 & \colhead{(J2000)} & \colhead{(J2000)} & & & &
}
\startdata
C32 & 22:17:25.63 & 0:16:12.9 & 3.301 & 3.290 & 23.64 & $>29.2$ \\
C30 & 22:17:19.29 & 0:15:45.0 & 3.104 & 3.097 & 23.70 & $>27.5$ \\
C47 & 22:17:20.24 & 0:17:32.5 & 3.075 & 3.065 & 23.78 & $>25.7$ \\
D3 & 22:17:32.40 & 0:11:33.6 & 3.086 & 3.077 & 23.92 & $>22.4$ \\
C35 & 22:17:20.23 & 0:16:52.5 & 3.103 & 3.098 & 24.06 & $>19.7$ \\
C24 & 22:17:18.94 & 0:14:45.4 & 3.102 & 3.091 & 24.19 & $>17.6$ \\
C11 & 22:17:25.68 & 0:12:35.4 & 3.109 & 3.099 & 24.21 & $>17.3$ \\
C12 & 22:17:35.29 & 0:12:47.9 & 3.118 & 3.106 & 24.22 & $>17.0$ \\
C4 & 22:17:38.91 & 0:11:02.0 & 3.076 & \nodata & 24.28 & $>16.2$ \\
MD23 & 22:17:28.01 & 0:14:29.6 & 3.092 & 3.082 & 24.34 & $>15.2$ \\
MD14 & 22:17:37.91 & 0:13:43.9 & \nodata & 3.094 & 24.49 & $>13.4$ \\
M10 & 22:17:26.80 & 0:12:21.3 & 3.103 & 3.095 & 24.50 & $>13.1$ \\
C48 & 22:17:18.58 & 0:18:16.7 & 3.090 & 3.079 & 24.57 & $>12.3$ \\
M28 & 22:17:31.66 & 0:16:58.0 & 3.094 & 3.088 & 24.75 & $>10.4$ \\
C28 & 22:17:21.13 & 0:15:27.7 & 3.076 & \nodata & 24.87 & $>9.3$ \\
C50 & 22:17:37.68 & 0:18:21.2 & \nodata & 3.086 & 25.01 & $>8.2$ \\
C26 & 22:17:39.54 & 0:15:15.6 & 3.178 & \nodata & 25.01 & $>8.2$ \\
C15 & 22:17:26.13 & 0:12:55.4 & 3.094 & \nodata & 25.02 & $>8.2$ \\
C39 & 22:17:20.99 & 0:17:09.5 & 3.076 & \nodata & 25.04 & $>8.0$ \\
M14 & 22:17:39.05 & 0:13:30.1 & 3.091 & \nodata & 25.20 & $>6.9$ \\
\enddata
\tablenotetext{a}{\mbox{Ly$\alpha$} emission redshift.}
\tablenotetext{b}{Instellar absorption redshift.}
\tablenotetext{c}{\mbox{Lower-limit} for observed ratio of non-ionizing UV and LyC emission with no correction for IGM absorption.}
\end{deluxetable*}

\clearpage

\begin{deluxetable*}{cccccccc}
\tablewidth{0pt} 
\footnotesize
\tablecaption{Photometry for LAEs without NB3640 detections. \label{t:laes2}}
\tablehead{
\colhead{ID} & \colhead{RA} & \colhead{Dec} & \colhead{4980} &
\colhead{$BV-$NB4980} & \colhead{$EW_0$ \tablenotemark{a}} & \colhead{$R$} & \colhead{$\frac{F_{UV}}{F_{LyC}}_{obs}$ \tablenotemark{b}} \\
 & \colhead{(J2000)} & \colhead{(J2000)} & & & \colhead{(\AA)} & &
}
\startdata
001\tablenotemark{c} & 22:17:32.40 & 0:11:34.1 & 23.00 & 0.74 & 21 & 23.92 & $>22.4$ \\
002\tablenotemark{d} & 22:17:38.90 & 0:11:01.8 & 23.15 & 1.41 & 61 & 24.28 & $>16.2$ \\
004\tablenotemark{e} & 22:17:28.01 & 0:14:30.0 & 23.40 & 1.06 & 36 & 24.34 & $>15.2$ \\
005 & 22:17:35.86 & 0:15:59.4 & 23.74 & 2.68 & 400 & 25.65 & $>4.6$ \\
006 & 22:17:24.80 & 0:11:16.8 & 23.76 & 1.69 &  90 & $>27$ & \nodata \\
007 & 22:17:27.78 & 0:17:36.9 & 23.84 & 1.96 & 131 & $>27$ & \nodata \\
008\tablenotemark{f} & 22:17:21.11 & 0:15:28.0 & 24.00 & 0.95 & 30 & 24.87 & $>9.3$ \\
009\tablenotemark{g} & 22:17:28.29 & 0:12:12.3 & 24.02 & 1.66 & 87 & 25.84 & $>3.8$ \\
011 & 22:17:33.85 & 0:12:14.9 & 24.15 & 1.54 &  73 & 26.07 & $>3.1$ \\
012\tablenotemark{h} & 22:17:31.69 & 0:16:57.6 & 24.23 & 0.97 &  31 & 24.75 & $>10.5$ \\
013 & 22:17:27.18 & 0:16:21.7 & 24.23 & 1.91 & 122 & 25.98 & $>3.4$ \\
014 & 22:17:19.25 & 0:14:50.9 & 24.27 & 1.42 &  62 & 25.82 & $>3.9$ \\
015 & 22:17:21.84 & 0:12:12.7 & 24.28 & 1.07 &  37 & $>27$ & \nodata \\
017 & 22:17:25.40 & 0:17:16.8 & 24.43 & 2.14 & 170 & 26.22 & $>2.7$ \\
020 & 22:17:37.33 & 0:16:31.4 & 24.55 & 1.44 &  64 & 25.45 & $>5.5$ \\
022 & 22:17:19.68 & 0:11:49.4 & 24.58 & 1.80 & 105 & 26.11 & $>3.0$ \\
023 & 22:17:31.73 & 0:16:06.9 & 24.61 & 0.88 &  27 & 24.91 & $>9.1$ \\
024 & 22:17:34.17 & 0:16:09.7 & 24.68 & 2.11 & 162 & 26.73 & $>1.7$ \\
026 & 22:17:18.96 & 0:12:00.8 & 24.70 & 1.63 &  83 & 26.59 & $>1.9$ \\
027 & 22:17:24.94 & 0:17:17.3 & 24.74 & 1.59 &  79 & 26.25 & $>2.6$ \\
029\tablenotemark{i} & 22:17:31.49 & 0:12:55.0 & 24.77 & 0.82 &  24 & 25.38 & $>5.9$ \\
030 & 22:17:21.75 & 0:11:38.8 & 24.81 & 1.34 &  55 & $>27$ & \nodata \\
031 & 22:17:33.63 & 0:17:15.1 & 24.83 & 1.67 &  88 & 26.37 & $>2.3$ \\
032 & 22:17:26.61 & 0:13:18.1 & 24.83 & 1.17 &  43 & 26.61 & $>1.9$ \\
033 & 22:17:37.50 & 0:14:08.3 & 24.85 & 1.58 &  78 & 26.62 & $>1.9$ \\
035 & 22:17:27.03 & 0:13:13.2 & 24.87 & 0.90 &  28 & $>27$ & \nodata \\
036 & 22:17:22.25 & 0:11:55.1 & 24.89 & 1.85 & 113 & $>27$ & \nodata \\
037 & 22:17:20.96 & 0:18:07.3 & 24.89 & 0.97 &  31 & 25.61 & $>4.7$ \\
040 & 22:17:31.93 & 0:13:08.5 & 24.92 & 1.96 & 131 & $>27$ & \nodata \\
042 & 22:17:21.50 & 0:17:04.7 & 24.93 & 1.04 &  35 & 25.50 & $>5.2$ \\
043 & 22:17:21.65 & 0:12:23.4 & 24.98 & 1.30 &  52 & 26.24 & $>2.7$ \\
044 & 22:17:36.41 & 0:12:51.0 & 24.99 & 1.36 &  57 & $>27$ & \nodata \\
045 & 22:17:35.97 & 0:16:30.2 & 25.03 & 1.45 &  65 & $>27$ & \nodata \\
047 & 22:17:36.05 & 0:15:06.9 & 25.04 & 1.96 & 131 & $>27$ & \nodata \\
049 & 22:17:39.29 & 0:16:10.5 & 25.06 & 2.17 & 177 & $>27$ & \nodata \\
050 & 22:17:24.56 & 0:15:56.8 & 25.08 & $>2.42$ & $>258$ & $>27$ & \nodata \\
052 & 22:17:36.84 & 0:13:17.2 & 25.13 & 1.04 &  35 & 26.63 & $>1.8$ \\
054 & 22:17:39.05 & 0:11:33.9 & 25.18 & 1.85 & 113 & 26.25 & $>2.6$ \\
055 & 22:17:35.80 & 0:11:50.0 & 25.20 & 1.81 & 107 & 26.67 & $>1.8$ \\
056 & 22:17:22.42 & 0:17:20.7 & 25.21 & 1.75 &  98 & 26.86 & $>1.5$ \\
057 & 22:17:25.40 & 0:10:58.3 & 25.23 & 1.07 &  37 & 26.84 & $>1.5$ \\
058 & 22:17:19.61 & 0:15:38.4 & 25.27 & 1.44 &  64 & $>27$ & \nodata \\
059 & 22:17:24.98 & 0:12:30.0 & 25.29 & 0.85 &  25 & 25.31 & $>6.2$ \\
060 & 22:17:28.19 & 0:11:17.1 & 25.30 & 1.16 &  42 & 26.61 & $>1.9$ \\
061 & 22:17:34.10 & 0:15:40.2 & 25.31 & $>2.19$ & $>182$ & $>27$ & \nodata \\
062 & 22:17:22.87 & 0:14:41.7 & 25.31 & 1.26 &  49 & 26.53 & $>2.0$ \\
063 & 22:17:23.32 & 0:15:52.9 & 25.31 & 1.63 &  83 & 26.55 & $>2.0$ \\
065 & 22:17:28.15 & 0:14:36.4 & 25.38 & $>2.12$ & $>165$ & 26.91 & $>1.4$ \\
066 & 22:17:20.86 & 0:15:11.8 & 25.41 & 0.73 &  20 & 26.64 & $>1.8$ \\
067 & 22:17:36.26 & 0:13:11.7 & 25.42 & 1.13 &  40 & 26.40 & $>2.3$ \\
068 & 22:17:18.37 & 0:17:26.1 & 25.44 & 1.28 &  51 & $>27$ & \nodata \\
070 & 22:17:39.28 & 0:14:00.2 & 25.45 & 0.78 &  22 & 25.98 & $>3.4$ \\
071 & 22:17:21.61 & 0:12:20.5 & 25.48 & 1.23 &  47 & 26.77 & $>1.6$ \\
072 & 22:17:31.24 & 0:17:32.1 & 25.48 & 1.78 & 102 & 27.00 & $>1.3$ \\
073 & 22:17:39.12 & 0:17:11.7 & 25.50 & 1.98 & 135 & 26.35 & $>2.4$ \\
075 & 22:17:22.97 & 0:11:25.8 & 25.51 & $>1.99$ & $>137$ & $>27$ & \nodata \\
076 & 22:17:20.67 & 0:15:13.2 & 25.54 & 1.13 &  40 & $>27$ & \nodata \\
078 & 22:17:37.68 & 0:16:48.3 & 25.56 & 1.21 &  46 & 25.95 & $>3.5$ \\
079 & 22:17:34.68 & 0:11:10.5 & 25.56 & 2.19 & 182 & $>27$ & \nodata \\
080 & 22:17:35.95 & 0:13:43.3 & 25.58 & 1.33 &  54 & $>27$ & \nodata \\
082 & 22:17:35.44 & 0:16:47.6 & 25.60 & 1.53 &  72 & $>27$ & \nodata \\
085 & 22:17:30.86 & 0:14:38.2 & 25.65 & 1.16 &  42 & 26.94 & $>1.4$ \\
086 & 22:17:28.42 & 0:13:42.8 & 25.65 & $>1.85$ & $>113$ & $>27$ & \nodata \\
088 & 22:17:38.45 & 0:13:18.3 & 25.74 & 1.25 &  48 & $>27$ & \nodata \\
089 & 22:17:38.54 & 0:15:22.5 & 25.74 & $>1.76$ & $>100$ & 26.59 & $>1.9$ \\
090 & 22:17:18.25 & 0:14:06.4 & 25.75 & 1.30 &  52 & $>27$ & \nodata \\
091 & 22:17:36.14 & 0:15:40.7 & 25.78 & $>1.72$ &  $>94$ & $>27$ & \nodata \\
092 & 22:17:23.97 & 0:15:27.8 & 25.79 & $>1.71$ & $>93$ & $>27$ & \nodata \\
093 & 22:17:27.48 & 0:13:57.5 & 25.80 & 0.78 &  22 & 26.35 & $>2.4$ \\
094 & 22:17:39.14 & 0:17:00.6 & 25.82 & $>1.68$ &  $>89$ & $>27$ & \nodata \\
095 & 22:17:37.19 & 0:13:28.0 & 25.82 & 1.13 &  40 & 26.17 & $>2.8$ \\
097 & 22:17:27.11 & 0:14:08.7 & 25.85 & 0.98 &  32 & $>27$ & \nodata \\
098 & 22:17:24.01 & 0:13:19.5 & 25.86 & $>1.64$ &  $>84$ & $>27$ & \nodata \\
099 & 22:17:36.46 & 0:13:00.3 & 25.87 & $>1.63$ &  $>83$ & $>27$ & \nodata \\
100 & 22:17:30.61 & 0:18:11.6 & 25.89 & 1.26 &  49 & $>27$ & \nodata \\
103 & 22:17:19.40 & 0:15:26.1 & 25.93 & $>1.57$ &  $>76$ & $>27$ & \nodata \\
104 & 22:17:37.66 & 0:12:55.5 & 25.94 & $>1.56$ & $>75$ & $>27$ & \nodata \\
105 & 22:17:35.46 & 0:12:23.9 & 25.94 & 0.81 &  24 & 26.81 & $>1.6$ \\
106 & 22:17:22.86 & 0:17:57.8 & 25.96 & 1.54 &  73 & 26.16 & $>2.9$ \\
107 & 22:17:20.96 & 0:14:46.7 & 25.97 & 0.76 &  22 & 26.57 & $>2.0$ \\
108 & 22:17:24.78 & 0:17:40.4 & 25.97 & 1.02 &  34 & 26.84 & $>1.5$ \\
109 & 22:17:23.98 & 0:17:57.8 & 25.99 & 0.77 &  22 & 26.77 & $>1.6$ \\
110 & 22:17:19.50 & 0:15:57.6 & 25.99 & $>1.51$ & $>70$ & $>27$ & \nodata \\
111 & 22:17:31.14 & 0:16:42.9 & 26.07 & $>1.44$ & $>64$ & $>27$ & \nodata \\
112 & 22:17:32.72 & 0:15:54.2 & 26.07 & $>1.44$ & $>64$ & $>27$ & \nodata \\
113 & 22:17:24.80 & 0:13:26.9 & 26.07 & $>1.44$ & $>64$ & $>27$ & \nodata \\
114 & 22:17:34.50 & 0:14:20.0 & 26.08 & $>1.44$ & $>64$ & $>27$ & \nodata \\
115 & 22:17:33.46 & 0:17:01.2 & 26.09 & $>1.43$ & $>63$ & $>27$ & \nodata \\
116 & 22:17:28.00 & 0:12:14.2 & 26.10 & $>1.43$ & $>63$ & $>27$ & \nodata \\
117 & 22:17:39.08 & 0:12:01.9 & 26.11 & $>1.42$ & $>62$ & $>27$ & \nodata \\
119 & 22:17:25.63 & 0:12:47.8 & 26.18 & $>1.38$ & $>59$ & $>27$ & \nodata \\
120 & 22:17:26.76 & 0:10:59.8 & 26.25 & $>1.35$ & $>56$ & $>27$ & \nodata \\
121 & 22:17:26.44 & 0:15:27.5 & 26.30 & $>1.33$ & $>54$ & $>27$ & \nodata \\
122 & 22:17:38.19 & 0:14:03.7 & 26.30 & $>1.32$ & $>54$ & $>27$ & \nodata \\
123 & 22:17:35.06 & 0:17:26.0 & 26.31 & $>1.32$ & $>54$ & $>27$ & \nodata \\
124 & 22:17:22.80 & 0:17:48.7 & 26.32 & $>1.32$ & $>54$ & $>27$ & \nodata \\
125 & 22:17:38.02 & 0:14:03.6 & 26.32 & $>1.31$ & $>53$ & 26.79 & $>1.6$ \\
126 & 22:17:19.53 & 0:16:48.2 & 26.33 & $>1.31$ & $>53$ & $>27$ & \nodata \\
127 & 22:17:36.91 & 0:11:27.1 & 26.36 & $>1.29$ & $>51$ & $>27$ & \nodata \\
128 & 22:17:23.43 & 0:16:07.4 & 26.48 & $>1.23$ & $>47$ & $>27$ & \nodata \\
129 & 22:17:22.28 & 0:10:57.9 & 26.49 & $>1.23$ & $>47$ & $>27$ & \nodata \\
130 & 22:17:32.84 & 0:16:48.8 & 26.49 & $>1.23$ & $>47$ & $>27$ & \nodata \\
\enddata
\tablenotetext{a}{\mbox{Ly$\alpha$} rest equivalent width estimated from $BV-$NB4980 color.}
\tablenotetext{b}{\mbox{Lower-limit} for observed ratio of non-ionizing UV and LyC emission with no correction for IGM absorption.}
\tablenotetext{c}{\mbox{LBG} D3, $z_{em} = 3.086$.}
\tablenotetext{d}{\mbox{LBG} C4, $z_{em} = 3.076$.}
\tablenotetext{e}{\mbox{LBG} MD23, $z_{em} = 3.092$.}
\tablenotetext{f}{\mbox{LBG} C28, $z_{em} = 3.076$.}
\tablenotetext{g}{\mbox{LBG} candidate C9.}
\tablenotetext{h}{\mbox{LBG} M28, $z_{em} = 3.094$.}
\tablenotetext{i}{\mbox{LBG} candidate M13.}
\end{deluxetable*}

\clearpage

\end{document}